\documentclass{amsart}

\usepackage[foot]{amsaddr}                  
\usepackage{amsmath}                        
\usepackage{amssymb}                        
\usepackage{dsfont} 	                    
\usepackage{bm}                             
\usepackage{mathtools}                      
\usepackage[hypertexnames=false]{hyperref}  
\usepackage[scientific-notation=true]{siunitx}
\usepackage{blkarray}
\usepackage{enumerate}
\usepackage{centernot}
\usepackage{paralist}
\usepackage{stackrel}
\usepackage[table]{xcolor}
\usepackage[colorinlistoftodos,prependcaption,textsize=tiny]{todonotes} 
\usepackage{pagenote}
\usepackage{setspace}
\usepackage[english]{babel}
\usepackage{subfig}
\usepackage{graphicx}
\usepackage{subfig}
\usepackage{dirtytalk}
\usepackage{placeins}

\usepackage[authoryear]{natbib}

\usepackage{algorithm}
\usepackage{algpseudocode}

\usepackage{xcolor}
\usepackage{listings}
\usepackage{color}

\definecolor{mycustomcolor}{rgb}{1, 0, 0} 

\usepackage{amsthm}
\theoremstyle{plain}
\theoremstyle{definition}

\theoremstyle{remark}

\newcommand{\LA}{\mathbf{\Lambda}}

\newcommand{\PS}{\mathbf{\Psi}}

\newcommand{\Covmat}{\boldsymbol{\Sigma}}

\newcommand{\Cormat}{\mathbf{R}}

\newcommand{\y}{\mathbf{y}}

\newcommand{\X}{\mathbf{X}}

\newcommand{\K}{\mathbf{K}}

\newcommand{\Z}{\mathbf{Z}}

\newcommand{\C}{\mathbf{R}}
\newcommand{\D}{\mathbf{D}}
\newcommand{\s}{\mathbf{s}}

\newcommand{\I}{\mathbf{I}}

\makeatletter
\def\squarecorner#1{
	\pgf@x=\the\wd\pgfnodeparttextbox%
	\pgfmathsetlength\pgf@xc{\pgfkeysvalueof{/pgf/inner xsep}}%
	\advance\pgf@x by 2\pgf@xc%
	\pgfmathsetlength\pgf@xb{\pgfkeysvalueof{/pgf/minimum width}}%
	\ifdim\pgf@x<\pgf@xb%
	\pgf@x=\pgf@xb%
	\fi%
	\pgf@y=\ht\pgfnodeparttextbox%
	\advance\pgf@y by\dp\pgfnodeparttextbox%
	\pgfmathsetlength\pgf@yc{\pgfkeysvalueof{/pgf/inner ysep}}%
	\advance\pgf@y by 2\pgf@yc%
	\pgfmathsetlength\pgf@yb{\pgfkeysvalueof{/pgf/minimum height}}%
	\ifdim\pgf@y<\pgf@yb%
	\pgf@y=\pgf@yb%
	\fi%
	\ifdim\pgf@x<\pgf@y%
	\pgf@x=\pgf@y%
	\else
	\pgf@y=\pgf@x%
	\fi
	\pgf@x=#1.5\pgf@x%
	\advance\pgf@x by.5\wd\pgfnodeparttextbox%
	\pgfmathsetlength\pgf@xa{\pgfkeysvalueof{/pgf/outer xsep}}%
	\advance\pgf@x by#1\pgf@xa%
	\pgf@y=#1.5\pgf@y%
	\advance\pgf@y by-.5\dp\pgfnodeparttextbox%
	\advance\pgf@y by.5\ht\pgfnodeparttextbox%
	\pgfmathsetlength\pgf@ya{\pgfkeysvalueof{/pgf/outer ysep}}%
	\advance\pgf@y by#1\pgf@ya%
}
\makeatother

\pgfdeclareshape{square}{
	\savedanchor\northeast{\squarecorner{}}
	\savedanchor\southwest{\squarecorner{-}}
	
	\foreach \x in {east,west} \foreach \y in {north,mid,base,south} {
		\inheritanchor[from=rectangle]{\y\space\x}
	}
	\foreach \x in {east,west,north,mid,base,south,center,text} {
		\inheritanchor[from=rectangle]{\x}
	}
	\inheritanchorborder[from=rectangle]
	\inheritbackgroundpath[from=rectangle]
}

\def\ImportFromMnSymbol#1{%
	\DeclareFontFamily{U} {MnSymbol#1}{}
	\DeclareFontShape{U}{MnSymbol#1}{m}{n}{
		<-6> MnSymbol#15
		<6-7> MnSymbol#16
		<7-8> MnSymbol#17
		<8-9> MnSymbol#18
		<9-10> MnSymbol#19
		<10-12> MnSymbol#110
		<12-> MnSymbol#112}{}
	\DeclareFontShape{U}{MnSymbol#1}{b}{n}{
		<-6> MnSymbol#1-Bold5
		<6-7> MnSymbol#1-Bold6
		<7-8> MnSymbol#1-Bold7
		<8-9> MnSymbol#1-Bold8
		<9-10> MnSymbol#1-Bold9
		<10-12> MnSymbol#1-Bold10
		<12-> MnSymbol#1-Bold12}{}
	\DeclareSymbolFont{MnSy#1} {U} {MnSymbol#1}{m}{n}
}
\newcommand\DeclareMnSymbol[4]{\DeclareMathSymbol{#1}{#2}{MnSy#3}{#4}}
\ImportFromMnSymbol{A}
\DeclareMnSymbol{\ConIndepNat}{\mathrel}{A}{225}

\makeatletter
\newcommand{\addresseshere}{%
  \enddoc@text\let\enddoc@text\relax
}

\definecolor{lightgrey}{rgb}{0.9,0.9,0.9}
\definecolor{darkgreen}{rgb}{0,0.6,0}
\calclayout

\usepackage{enumitem}
\usepackage{tikz}
\tikzstyle{node}=[very thick, circle, draw=black, minimum size=22, inner sep=0.8, outer sep=0.6]

\usetikzlibrary{calc}
\usetikzlibrary{arrows.meta}
\usetikzlibrary{chains}
\usetikzlibrary{arrows, arrows.meta}

\DeclareRobustCommand{\rvdots}{%
	\vbox{
		\baselineskip4\p@\lineskiplimit\z@
		\kern-\p@
		\hbox{.}\hbox{.}\hbox{.}
}}

\begin{document}
\sloppy

\title[Kernel-based models for G$\times$E$\times$M interactions]
{REML implementations of kernel-based genomic prediction models for genotype $\times$ environment $\times$ management interactions}

\author{Killian A.C.\ Melsen$^{1*}$}
\author{Salvador A. Gezan$^2$}
\author{Daniel J.\ Tolhurst$^3$}
\author{Fred A.\ van Eeuwijk$^1$}
\author{Carel F.W.\ Peeters$^1$}

\address{$^1$Mathematical \& Statistical Methods group - Biometris, Wageningen University \& Research, PO Box 16, 6700 AA, Wageningen, The Netherlands}
\address{$^2$VSN International, 2 Amberside House, Wood Lane, HP2 4TP, Hemel Hempstead, United Kingdom}
\address{$^3$The Roslin Institute and Royal (Dick) School of Veterinary Science, University of Edinburgh, Easter Bush, Midlothian, EH25 9RG, United Kingdom}

\email{$^*$killian.melsen@wur.nl}


\maketitle
\thispagestyle{headings}
\newpage
\begin{abstract}
	\label{abstract}
	High-throughput pheno-, geno-, and envirotyping allows characterization of plant genotypes and the trials they are evaluated in, producing different types of data.
	These different data modalities can be integrated into statistical or machine learning models for genomic prediction in several ways.
	One common approach within the analysis of multi-environment trial data in plant breeding is to create linear or nonlinear kernels which are subsequently used in linear mixed models for genotype by environment interactions.
	Current implementations of these kernel-based mixed models present several opportunities in terms of methodological extensions.
	Here we show how these models can be easily implemented in standard software, allowing direct restricted maximum likelihood estimation of all parameters.
	We also extend the models by combining the kernels with unstructured covariance matrices for three-way interactions in genotype by environment by management datasets, while simultaneously allowing for environment-specific genetic variances and multiple bandwidth parameters.
	We show how the models incorporating Gaussian kernels, heterogeneous variances, and management-specific bandwidths maximize the amount of genetic variance captured by environmental covariables, following a biologically intuitive pattern.
	Furthermore, we show that these models perform best in sparse testing prediction settings using two real-word datasets on wheat and maize.
	We discuss opportunities regarding models with multiple kernels or kernels obtained after feature selection, as well as the similarities to models regressing phenotypes on latent and observed environmental covariables.
	Finally, we discuss the flexibility provided by our implementation in terms of modeling complex plant breeding datasets, allowing for straightforward integration of phenomics, enviromics, and genomics.
\end{abstract}
\newpage
\section*{Summary}
We have developed a flexible statistical approach for the integration of genomic, environmental, and phenotypic data for genomic prediction in plant breeding.
Our implementation makes it easy to fit linear mixed models for genotype by environment by management interactions using similarity matrices for genotypes and environments based on genomic and weather data.
The results show that our approach results in a larger part of the genetic variation being captured by environmental covariables as well as a higher prediction accuracy.
The {\ttfamily R} package we provide allows plant breeders to easily fit these complex models using software they are already familiar with.
\newpage


\section{Introduction}
Genomic prediction has become widespread in the field of plant breeding.
In its most basic form, genomic prediction attempts to model genetic estimated breeding values (GEBVs) for quantitative traits as a simple function of genomic single nucleotide polymorphism (SNP) marker scores.
The practice of genomic prediction has had a significant impact on the field of plant breeding since its introduction \citep{Meuwissenetal_2001, Bernardo_1994}.
It allows for more efficient allocation of resources within a breeding program and an increased rate of genetic gain by modifying several parameters of the breeder's equation \citep{LynchandWalsh_1998}.
Genomic selection, i.e., selection based on GEBVs rather than phenotypes, typically allows for higher selection intensities due to larger populations, higher selection accuracies, and possibly shorter generation intervals \citep{Crossaetal_2017}.
A wide variety of models have been used to obtain GEBVs \citep{Perez-EncisoandZingaretti_2019, LiandSillanpaa_2012}.
While several machine and multimodal deep learning approaches have recently emerged as methods for genomic prediction \citep{Potzeetal_2026, Crossaetal_2025}, linear mixed models (LMM) remain among the most widely used models due to their ability to model dependencies in the data through random effects, interpretability, and ability to partition observed variance.
Instead of direct random regression on genomic SNP scores, LMMs typically include a random effect which is assumed to have a covariance matrix that is proportional to a relationship matrix of genotypes based on pedigree information or the SNP scores themselves \citep{VanRaden_2008}.
The LMMs can then be used to obtain predicted breeding values in the form of best linear unbiased predictions (BLUPs) for genotypes, regardless of whether these genotypes were phenotyped in field trials.
The models also readily provide estimates of heritabilities which can be used to estimate other quantities like response to selection.

While the LMMs can be used to predict GEBVs within a single trial, in which case it is typically referred to as GBLUP, they are also commonly used to analyze multi-environment trial (MET) data.
Here, genotypes are evaluated in several locations and years that together represent a sample of environments from the target population of environments (TPE), i.e., the set of environments targeted by plant breeders \citep{Bancicetal_2024, Piepho&Williams_2024, Cooperetal_2022, Cooperetal_2002}.
Within environments, co-located trials can be used to evaluate genotypes under a number of different management practices that may be used on farms.
Examples of such management practices are rain-fed versus irrigated fields, fields with different nitrogen fertilization rates, or different planting densities.
See \citep{Wangetal_2025, Krauseetal_2019, BandeiraeSousaetal_2017, milletetal_2016} for examples of datasets containing such MET trial designs.
Data obtained from these multi-environment, multi-management trials can be analyzed using LMMs.
A commonly used approach is to consider the phenotypes of genotypes under various combinations of managements and environments as different traits in a classical multivariate quantitative genetics framework \citep{FalconerandMackay_1996}.
This involves considering managements as a separate categorical variable or factor that is nested within environments.
While there are arguments to be made for the idea that management practices should be considered part of the environment, there are a number of advantages to the inclusion of management as an aditional categorical variable if the MET structure allows.
First of all, it allows for the estimation of genetic correlations between management conditions from the phenotypic data, resulting in appropriate borrowing of information.
Furthermore, it allows for any developed modeling approaches to be used directly for actual multi-trait data, such as the joint analysis of grain yield and protein content, or a trait of interest and several features derived from high-throughput phenotyping.

One key aspect of the approach described above is that all genetic covariances between combinations of managements and environments must typically be estimated from the phenotypic data itself.
While the assumption of a simple covariance structure with few parameters can simplify this task, complex, three-way genotype by environment by management (G$\times$E$\times$M) interactions typically require more flexible structures \citep{Malosettietal_2013}.
However, the large number of parameters underlying these flexible covariance structures can be hard to estimate.
This is especially true if genotypes are only evaluated in small subsets of environments due to an MET design known as \textit{sparse testing} which allows for larger samples of the TPE without increasing the total number of plots \citep{Jarquinetal_2020}.

The increasing use of high-throughput technologies in recent years has provided a possible solution to the problems associated with the estimation of the large numbers of covariance parameters typically required to model MET datasets.
The actual number of parameters that need to be estimated can be greatly reduced by replacing unstructured covariance matrices by known similarity matrices based on features obtained from high-throughput technologies.
High-throughput methods for characterizing genotypes using SNP markers have been commonly available for a large number of agricultural crops in the past few years.
As a result, covariances between genotypes are typically modeled using a product of the previously mentioned relationship matrices and a scalar genetic variance.
Recently, cheap sensor technology has made high-throughput characterization of environments possible as well.
Weather stations now collect high-resolution longitudinal data on environmental covariables such as temperature, soil water potential, humidity, and solar radiation for every environment within an MET breeding program \citep{milletetal_2019}.
Furthermore, satellite data like those obtained from the NASA POWER (National Aeronautics and Space Administration, Prediction Of Worldwide Energy Resources) project allow environmental characterization of any location on earth \citep{Monteiroetal_2017}.
Like for genotypes, these features can easily be used to construct relationship matrices of environments, also referred to as kernels.
There has been significant interest in using these types of \textit{enviromics} kernels to model G$\times$E interactions, resulting in several approaches \citep{Costa-Netoetal_2021a, Crossaetal_2021, jarquinetal_2014, Malosettietal_2004}.

Among these approaches, reaction norm models using kernels as substitutes for between-environment genetic covariance matrices have been a commonly used choice \citep{Perez-rodriguez&deloscampos_2022, jarquinetal_2014, deloscamposetal_2010}.
While the term kernel is used in several contexts in the literature, in the analysis of MET data using LMMs, a kernel typically refers to a positive definite function that produces a Gram matrix containing inner products of vectors $\mathbf{x}$ and $\mathbf{x}'$ after mapping the vectors from the original space $\mathcal{X}$ to some other space $\mathcal{V}$ \citep{Halperin_1962}.
One attractive feature of these kernel methods is the ability to obtain the Gram matrix without performing the explicit mapping $\mathcal{X}\rightarrow\mathcal{V}$.
For environments, these vectors contain environmental features derived from covariables that aim to characterize the environment from a crop growth perspective.

Kernel-based LMMs avoid the need to estimate covariance parameters by essentially regressing the phenotypes on products of environmental features and SNP marker scores.
They achieve this by assuming that the covariance matrix of the random effect can be expressed as a Kronecker or Hadamard product of appropriately sized and known kernels \citep{martinietal_2020, jarquinetal_2014}.
These kernels can be linear, i.e., $\mathcal{X}=\mathcal{V}$, or nonlinear and have been used in a wide variety of models, ranging from support vector machines and other machine learning approaches to the LMMs mentioned above \citep{Costa-Netoetal_2021a, KeerthiandLin_2003}.
While the use of both linear and nonlinear kernels avoids the need to estimate many covariance parameters, which is especially useful in sparse testing scenarios, an additional benefit of the nonlinear kernels in particular is their flexibility due to the introduction of one or more additional parameters.
A commonly used example of such nonlinear kernels is the Gaussian or radial basis function (RBF) kernel which introduces nonlinearity and flexibility through a single parameter often referred to as the bandwidth or scale parameter \citep{Costa-Netoetal_2021a, Montesinos-Lópezetal_2021}.
This flexibility is a result of the so-called {\it kernel trick} that allows for nonlinearity by mapping the original features, e.g., SNP marker scores or environmental features, to a higher-dimensional space \citep{Montesinos-Lopezetal_2022, Scholkopf_2000}.
Note that the LMMs relying on Gaussian kernels described above are closely related to the machine learning approaches known as Gaussian process models often used in the context of geostatistics and other fields.
\citet{Bonillaetal_2007} provide details on a multi-task Gaussian process model, which is essentially equivalent to the common multi-environment LMM as noted by \citet{Cuevasetal_2017}, as well as related work.
As mentioned, kernels are also widely used in other machine learning approaches leading to strong similarities between LMMs and approaches such as least squares support vector machine (LS-SVM) regression and reproducing kernel Hilbert spaces (RKHS) regression as noted by \citet{Piepho_2009} and shown by \citet{Suykensetal_2002} and \citet{GianolaandvanKaam_2008}.

While the mentioned reaction norm models often work well, there are a few opportunities for improving existing implementations.
For example, reaction norm models typically estimate a single genetic variance for all environments, thus assuming the same heritability and rate of shrinkage for all environments if a single residual variance is used.
Furthermore, selection of appropriate bandwidth parameters for Gaussian kernels is often computationally expensive \citep{Caamal-Patetal_2021}.
Strategies for bandwidth selection typically rely on cross-validation, Bayesian methods, or a technique known as kernel averaging \citep{Perez-Elizaldeetal_2015, deloscamposetal_2010, GianolaandvanKaam_2008}.
\citet{Endelman_2011} developed the {\ttfamily R}-package {\ttfamily rrBLUP} that allows convenient restricted maximum likelihood (REML) estimation of the bandwidth, but only in the context of LMMs with a single genetic variance component, i.e., no G$\times$E or G$\times$E$\times$M interaction models, and it still relies on a grid of discrete values.
We note that direct REML estimation of the bandwidth is possible within the context of spatial modeling, but only in a limited set of applications.
One example involves GBLUP-like models where SNP marker scores are treated as coordinates in a high-dimensional marker space and a single bandwidth transforms high-dimensional distances into correlations \citep{Piepho_2009}.
However, fitting a Gaussian kernel for environments as part of a Kronecker product with a kinship matrix is not possible within this framework.
While other hyperparameters such as the smoothing parameter of a penalized spline can be estimated in complex LMMs as a ratio of variance components \citep{Ruppertetal_2003}, we are not aware of any implementation of REML estimation of bandwidths within complex interaction models.
Finally, we note that a pure Kronecker covariance structure for three-way G$\times$E$\times$M interactions results in a fully separable structure.
While this may be considered a benefit as it allows for the separation of confounded managements and environments if a kernel is used for the latter, it also assumes that the G$\times$E structure does not vary across managements.
Relaxing this assumption by estimating management-specific kernels may improve model fit and prediction accuracy.

Here, we propose several extensions to the kernel-based LMMs.
We show how models using linear and Gaussian kernels combined with unstructured covariance matrices for complex G$\times$E$\times$M data can be fitted within a frequentist framework in a straightforward way using standard LMM software, allowing for direct REML estimation of all parameters.
In addition to REML estimation of all parameters, we allow for the estimation of heterogeneous genetic variances for kernel models resulting in environment-specific rates of shrinkage and heritabilities.
We also extend the Gaussian kernel models by estimating management-specific bandwidths, allowing G$\times$E to vary across managements.
We apply the different models to two real-world G$\times$E$\times$M maize and wheat datasets in a sparse testing context to (1) investigate the benefit of using environmental covariables to model G$\times$E when connectivity between environments is low, (2) show the impact of the additional flexibility of the Gaussian kernel, (3) show the benefit of allowing for heterogeneous genetic variances, and (4) evaluate the potential advantage of management-specific bandwidths.
Note that while we opted for a weighted two-stage approach to illustrate the proposed models, all covariance structures discussed in this paper can be fitted within a single-stage framework as well \citep{Damesaetal_2017}.
From an interpretability point of view, we realize that while the environmental features underlying the kernels aim to characterize environments, they cannot perfectly do so.
We therefore make use of the LMM's ability to partition genetic variance into kernel and lack of fit components.
By doing so, we show that while not all G$\times$E can be predicted from environmental features, a large proportion of the total variance is captured by the kernels regardless of the management.
We also show that when G$\times$M is negligible, but G$\times$E$\times$M is not, management-specific bandwidths increase prediction accuracy and result in kernels that imply different levels of G$\times$E across managements, matching what would be expected based on biological knowledge.
We finally provide easy-to-use code in the form of an {\ttfamily R}-package.

The remainder of this paper is organized as follows.
In Section \ref{sec:MM} we provide some details on the notation used (\ref{sec:MM:notation}), a description of the benchmark and kernel-based models (\ref{sec:models}), details on the variance partitioning analyses (\ref{sec:LOF_variance}), cross-validation setup and genomic prediction scenario considered (\ref{sec:MM:GP}), and software implementation (\ref{sec:MM:software}), as well as a description of the two datasets and any pre-processing of the data (\ref{sec:MM:BRIWECS} and \ref{sec:MM:DROPS}).
Section \ref{sec:RD} then provides results on these two datasets, showing differences between models in terms of variance partitioning and genomic prediction accuracy, as well as a discussion.
We conclude in Section \ref{sec:C}.

\section{Materials and methods}\label{sec:MM}
\subsection{Notation}\label{sec:MM:notation}
In terms of notation, we denote vectors and matrices with bold lower- and upper-case letters, respectively, while we use normal characters for scalars.
Covariance matrices are denoted using $\Covmat$ while we use $\C$ for correlation matrices.
We use subscripts $M$, $E$, and $G$ to denote covariance structures related to managements, environments, and genotypes, respectively.
For residual covariance matrices and variances we use a sub- or superscript $\epsilon$.
We use $\I_n$ for the identity matrix of size $n$, $\mathbf{J}_n$ for an n$\times$n matrix of ones, $\D$ for a distance matrix containing squared Euclidian distances, and $\K$ for a linear kernel, including the genomic relationship or kinship matrix $\K_G$.
The $\circ$ and $\otimes$ operators denote the Hadamard and Kronecker products.
For a matrix $\mathbf{A}=\left[a_{ij}\right]$ we use $\mathbf{B} = e^\mathbf{A}$ to define the matrix $\mathbf{B}=\left[e^{a_{ij}}\right]$.
We use $*$ to denote a special case of the Khatri-Rao or block Kronecker product so that for $\mathbf{A}*\mathbf{B}$ each scalar element of $\mathbf{A}\in\mathbb{R}^{p\times p}$ is multiplied by the corresponding square block of $\mathbf{B}\in\mathbb{R}^{pq\times pq}$.
Finally, let $\text{diag}\left(\mathbf{a}\right)$ denote the diagonal matrix with elements $a_i$ on the diagonal.

\subsection{Modeling approaches}\label{sec:models}
\begin{figure}
	\begin{tikzpicture}
		\node[inner sep=0pt] (pic) at (0, -1.3) {\includegraphics[width=\textwidth]{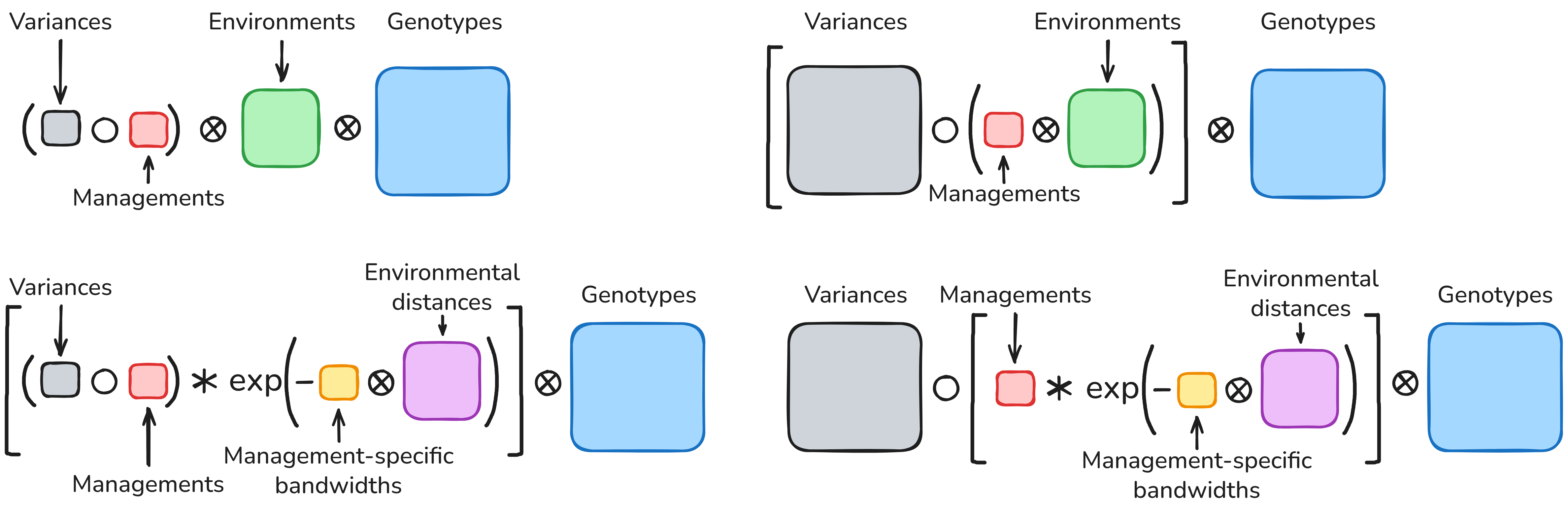}};
		\node (A) at (-7.2, 1.2) {\Large \textbf{a}};
		\node (B) at (-0.2, 1.2) {\Large \textbf{b}};
		\node (C) at (-7.2, -1.3) {\Large \textbf{c}};
		\node (D) at (-0.2, -1.3) {\Large \textbf{d}};
	\end{tikzpicture}
	\caption{Graphical representation of the different kernel-based G$\times$E$\times$M covariance structures.
		For the single bandwidth, single variance models (\textbf{a}), variances (gray) are first multiplied with the unstructured correlation matrix for managements before forming the Kronecker product with the kernel for environments.
		For the single bandwidth, multiple variance models (\textbf{b}), the unstructured correlation matrix for managements (red) is first combined with the kernel for environments (green) and expanded to all E$\times$M combinations before multiplication with the environment-specific variances.
		Panels \textbf{c} and \textbf{d} represent single and multiple variance versions of Gaussian kernel-based models with unique bandwidths for each management.
		The $*$ denotes a special case of the Khatri-Rao or block Kronecker product where each scalar value in the matrix for managements is multiplied with the corresponding square block of the exponential term \citep{Liu_1999}.
		The matrix of bandwidths is shown in orange and the squared Euclidian distance matrix of environments in purple.
		Note that the matrix of bandwidths is of the same dimensions as the correlation matrix of managements.
		All models involve a final Kronecker product with the kinship matrix (blue).
		Note that the kernel for the environments shown in green may be linear or nonlinear.
	\\[5pt]
	\textbf{Alt text:} A graphical overview showing the structures of the multiple variance and single variance kernel-based models.}
	\label{fig:models}
\end{figure}

We focus on multi-environment, multi-management trial data, i.e., data containing phenotypic records (grain yield) on multiple genotypes in multiple environments (location-year combinations), and under different management conditions within a given environment.
Grain yields under different managed treatments can thus be considered as different traits.
Let $\y \in \mathbb{R}^{pqr}$ be the vector of BLUEs (best linear unbiased estimates) for the G$\times$E$\times$M combinations where the $r$ genotypes are nested within the $q$ environments, which themselves are nested within $p$ managements:
\begin{equation*}
	\y = \begin{bmatrix}
		y_{111} & y_{112} & \dots & y_{ijk} & \dots & y_{pqr}
		\end{bmatrix}^\top \text{,}
\end{equation*}
where $i = 1,2\dots,p$, $j = 1,2\dots,q$, $k = 1,2\dots,r$, $pqr = n$, and $y_{ijk}$ is thus the phenotype for genotype $k$ in environment $j$ under management $i$.
We use the index $l = 1,2,\dots,pq$ to iterate over all E$\times$M combinations.
By employing a two-stage modeling approach where individual E$\times$M combinations are first analyzed to obtain the BLUEs in $\y$ and corresponding weights, the data can generally be analyzed in a second stage with LMMs of the following form:
\begin{equation}\label{eq:LMM}
	\mathbf{y} = \mathbf{X} \boldsymbol{\beta} + \mathbf{Z} \mathbf{u} + \boldsymbol{\epsilon} \text{,}
\end{equation}
where $\X \in \mathbb{R}^{n\times pq}$ is a fixed effect design matrix for the means of every E$\times$M combination and $\boldsymbol{\beta} \in \mathbb{R}^{pq}$ is the vector containing the estimates for these fixed effects.
We opted to take the environment and management main effects as fixed for all models since our interest within the sparse testing scenario is in the prediction of genotype-specific deviations for known environments and managements.
Note that fixed main effects are by no means required, however, and the environmental main effect may also be taken as random if prediction of new environments is the goal.
In that case, means for unseen environments can be predicted when values for the environmental covariables are provided \citep{Piepho&Williams_2024}.
The vector $\boldsymbol{\epsilon} \in \mathbb{R}^{n}$ contains the residuals which, for all models presented in this paper, are assumed to be independently distributed with mean $0$ and variance $\upsilon_{ijk} = 1 / \omega_{ijk}$ where $\omega_{ijk}$ is the $k$th diagonal element of $\boldsymbol{\Omega}_{ij}^{-1}$ which is the inverse covariance matrix of BLUEs obtained from the first-stage analysis of the plot-level data for management $i$ and environment $j$ \citep{Damesaetal_2017, Smithetal_2001}.
The random effect design matrix $\Z \in \mathbb{R}^{n \times n}$ links records to the nested G$\times$E$\times$M BLUPs in $\mathbf{u}$ which have mean zero.
It is the modeling of the covariance matrix of this nested random G$\times$E$\times$M effect that we are primarily concerned with.
In a general form this covariance matrix can be represented as
\begin{equation*}
	\mathbb{V}\text{ar}\left(\mathbf{u}\right) = \Covmat_M \otimes \Covmat_E \otimes \Covmat_G \text{,}
\end{equation*}
where $\Covmat_M$ is the between-managements genetic covariance matrix, $\Covmat_E$ is the between-environments genetic covariance matrix, and $\Covmat_G$ is a matrix containing additive genetic relationships between genotypes.
Note that $\Covmat_G$ may be a pedigree-based relationship matrix $\mathbf{A}_G$.
For the remainder of this paper, however, we will assume that $\Covmat_G$ is the SNP-based additive genomic relationship matrix $\K_G$, which is a linear kernel.
Table \ref{model_overview} provides an overview of the baseline and kernel-based models that are described in the remainder of this section in terms of the number of parameters underlying the covariance structures.
Figure \ref{fig:models} provides a graphical overview of the different kernel-based models.
We start by describing the baseline additive main effect and factor analytic models in Sections \ref{model:ME} and \ref{model:FA}, followed by six different kernel-based models of increasing complexity.

\subsubsection{Additive main effects (ADD)}\label{model:ME}
The baseline model contains a random effect for the genotypes only, thus treating all E$\times$M combinations as if they were perfectly genetically correlated.
As a result, the dimensions of $\Z$ and $\mathbf{u}$ for the ADD models differ slightly from those given for the general model (\ref{eq:LMM}).
For the main effects model, $\Z\in\mathbb{R}^{n \times r}$ links BLUEs in $\y$ to BLUPs in $\mathbf{u}\in\mathbb{R}^r$.
This means the model produces a single BLUP for every genotype and the covariance of $\mathbf{u}$ is modeled as
\begin{equation*}
	\mathbb{V}\text{ar}\left(\mathbf{u}\right) = \sigma^2 \K_G \text{,}
\end{equation*}
where $\sigma^2$ is the additive genetic variance.
Note that while this model does not contain any interaction effect, it does, like all other models, contain a fixed effect for all E$\times$M combinations.

\subsubsection{Order $m$ factor analytic (FA-$m$)}\label{model:FA}
The factor analytic (FA) models of different orders form the baseline interaction models.
FA models include G$\times$E$\times$M interactions by using the phenotypic data to estimate genetic correlations.
They are a more parsimonious alternative to fully unstructured models that typically cannot be fitted on datasets with more than a few environments \citep{butleretal_2023, tolhurstetal_2019}.
FA models usually collapse the management and environment factors into a new factor with $pq$ levels for three-way G$\times($E$\times$M) interactions. 
This factor is subsequently modeled using loadings $\LA \in \mathbb{R}^{pq \times m}$ and unique variances $\PS = \text{diag}\left(\boldsymbol{\psi}\right)$ for each E$\times$M combination, given a number of latent factors $m$:
\begin{equation*}
	\mathbb{V}\text{ar}\left(\mathbf{u}\right) = \left(\LA \LA^{\top} + \PS\right) \otimes \K_G \text{.}
\end{equation*}
Each E$\times$M combination is thus modeled using a unique additive genetic variance, and genomic predictions are obtained for all E$\times$M combinations for which phenotypic data are available.
Note that the loadings can be thought of as latent covariables.
Therefore, an important limitation of the factor analytic models implemented here is that extrapolation into new environments is not straightforward as loadings are estimated from the data \citep{Tolhurstetal_2022}.

\subsubsection{Single variance linear kernel (SV-LK)}
The standard multivariate reaction norm model essentially regresses phenotypes on the products of environmental features obtained from covariables and SNP marker scores using a single genetic variance for each level of management \citep{Perez-rodriguez&deloscampos_2022, jarquinetal_2014, deloscamposetal_2010}.
Literature on reaction norm models typically uses notation involving Hadamard products:
\begin{equation}\label{eq:hadamard}
	\mathbb{V}\text{ar}\left(\mathbf{u}\right) = \Z_M \Covmat_M \Z_M^\top \circ \Z_E \K_E \Z_E^{\top} \circ \Z_G \K_G \Z_G^{\top} \text{,}
\end{equation}
where $\Z_M \in \mathbb{R}^{n\times p}$, $\Z_E \in \mathbb{R}^{n\times q}$, and $\Z_G \in \mathbb{R}^{n\times r}$ are design matrices linking records to managements, environments, and genotypes, and $\Covmat_M$ is an unstructured genetic covariance matrix for the $p$ levels of managements.
Environmental features are used to construct the linear kernel for environments $\K_E$:
\begin{equation}\label{eq:linear_kernel}
	\K_E = \dfrac{\mathbf{W}^\top\mathbf{W}}{s - 1} \text{,}
\end{equation}
where $\mathbf{W}\in\mathbb{R}^{s \times q}$ is a column-wise scaled and centered matrix containing $s$ features obtained from environmental covariables for $q$ environments that have also been scaled and centered to ensure quantities measured in different units have comparable scales.
The linear kernel $\K_E$ is thus a correlation matrix, or Gram matrix containing inner products in the original $s$-dimensional space \citep{Halperin_1962}, meaning that the implicit mapping underlying the kernel is the identity map \citep{Costa-Netoetal_2021a}.
Additive genetic relationships between genotypes are given by $\K_G$, and as $\K_E$ is a correlation matrix, any estimated genetic variances can be interpreted as additive.
Note that in this formulation, unique additive genetic variances $\sigma^2_i$ are only estimated for every management level $i$, not every E$\times$M combination $l$, even if phenotypic data are available for all environments.
Also note that if the data are balanced, i.e., there are exactly $n=pqr$ records in $\y$, and ordered so that genotypes are nested within environments within managements, the notation using Hadamard products in (\ref{eq:hadamard}) can be freely exchanged with notation using Kronecker products \citep{martinietal_2020} (Figure \ref{fig:models}a):
\begin{equation}\label{svlk_kron}
	\mathbb{V}\text{ar}\left(\mathbf{u}\right) = \Covmat_M \otimes \K_E \otimes \K_G \text{.}
\end{equation}

\subsubsection{Multiple variance linear kernel (MV-LK)}
This model extends the previous covariance model (\ref{svlk_kron}) with unique variances for each combination of environment and management, allowing for environment-specific rates of shrinkage.
This results in the following covariance matrix for $\mathbf{u}$:
\begin{equation*}
	\mathbb{V}\text{ar}\left(\mathbf{u}\right) = \Big[\s\s^{\top} \circ \left(\C_M \otimes \K_E\right)\Big] \otimes \K_G \text{,}
\end{equation*}
where $\C_M$ is an unstructured correlation matrix of managements (Figure \ref{fig:models}b).
The linear kernel $\K_E$ and genomic relationship matrix $\K_G$ are as described before.
The column vector $\s$ of length $pq$ contains the additive genetic standard deviations for each E$\times$M combination.

\subsubsection{Single variance Gaussian kernel (SV-GK)}
\begin{figure}[h]
	\centering
	\includegraphics[width=\textwidth]{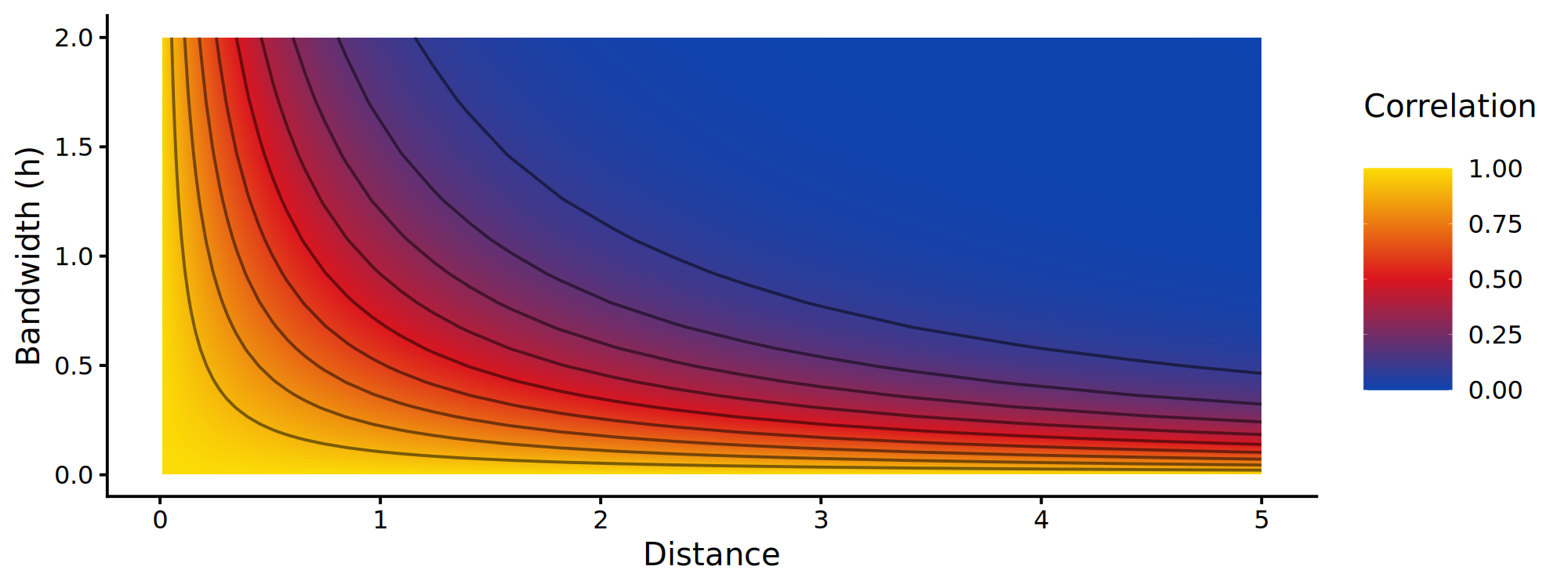}
	\caption{The Gaussian kernel nonlinearly transforms squared Euclidian distances between environments to genetic correlations, with the transformation depending on the bandwidth parameter $h\in \left(0,\infty\right)$. Genetic correlations tend to $0$ for large bandwidth values, regardless of the distance, while for small bandwidth values the opposite is true. Black contour lines are placed at $0.1$ intervals.
	\\[5pt]
	\textbf{Alt text:} A contour plot showing how the bandwidth and squared Euclidian distance affect the resulting correlation.}
	\label{fig:bandwidth}
\end{figure}

The next model is a REML implementation of the Gaussian kernel-based on environmental covariables for the environments.
While the previous two kernel-based models relied on a linear kernel, and thus implied an identity map resulting in a regression of phenotypes directly on the original environmental features, the kernel models described next use nonlinear, Gaussian kernels.
The Gaussian kernel $e^{-h\D}$, also referred to as the radial basis function (RBF) kernel, implies a mapping to infinite-dimensional space \citep{RingandEskofier_2016, Vertetal_2004}.
While regression on an infinite number of features is clearly impractical, the Gaussian kernel is readily implemented in an LMM through the \textit{kernel trick} by assuming the following covariance structure \citep{Montesinos-Lopezetal_2022, Piepho_2009, Scholkopf_2000}:
\begin{equation}\label{svgk}
	\mathbb{V}\text{ar}\left(\mathbf{u}\right) = \Covmat_M \otimes e^{-h\D_E} \otimes \K_G \text{,}
\end{equation}
where the symmetric matrix $\D_E$ contains squared Euclidian distances between environments (Figure \ref{fig:models}a).
The parameter $h \in \left(0,\infty\right)$ is the bandwidth determining the nonlinear transformation of squared Euclidian distances to correlations (Figure \ref{fig:bandwidth}).
The bandwidth thus introduces flexibility resulting in a model that is expected to fit the data better than the simpler linear kernel-based models \citep{Montesinos-Lópezetal_2021}.
Note again that $e^{-h\D_E}$ is a correlation matrix, meaning that any estimated genetic variances can be interpreted as additive.
The distances are computed by averaging the squared differences between features obtained from environmental covariables for environment $j$ and $j'$:
\begin{equation}\label{eq:gaussian_kernel}
	\D_{E_{jj'}} = \dfrac{1}{s}\sum_{b=1}^{s}\left(\mathbf{W}_{bj} - \mathbf{W}_{bj'}\right)^2 \text{,}
\end{equation}
where $\mathbf{W} \in \mathbb{R}^{s\times q}$ contains the $s$ features for $q$ environments and is row-wise centered and scaled.
Note that $\D_{E_{jj'}}=0$ if $j = j'$.

\subsubsection{Multiple variance Gaussian kernel (MV-GK)}
This model then extends (\ref{svgk}) with unique genetic variances for all combinations of environments and managements, combining the flexibility of the Gaussian kernel with environment-specific rates of shrinkage (Figure \ref{fig:models}b):
\begin{equation*}
	\mathbb{V}\text{ar}\left(\mathbf{u}\right) = \Big[\s\s^{\top} \circ \left(\C_M \otimes e^{-h\D_E}\right)\Big] \otimes \K_G \text{,}
\end{equation*}
where $\s$ again contains additive genetic standard deviations for all E$\times$M combinations.

\subsubsection{Multiple bandwidth Gaussian kernels (MB-SV-GK and MB-MV-GK)}
The final two models are versions of the single and multiple variance Gaussian kernels with management-specific bandwidths, relaxing the fully separable Kronecker structure of the previous kernel models.
For the single variance version (Figure \ref{fig:models}c, MB-SV-GK):
\begin{equation*}
	\mathbb{V}\text{ar}\left(\mathbf{u}\right) = \Big[\left(\Covmat_M \otimes \mathbf{J}_q\right) \circ e^{-\mathbf{H}\otimes\D_E}\Big] \otimes \K_G \text{,}
\end{equation*}
where $\mathbf{J}_q$ is a $q\times q$ matrix of ones and
\begin{equation*}
	\mathbf{H} = \begin{bmatrix}
		h_{1} & h_{12} & \dots & h_{1p} \\
		h_{21} & h_{2} & \dots & h_{2p} \\
		\vdots & \vdots & \ddots & \vdots \\
		h_{p1} & h_{p2} & \dots & h_{p}
	\end{bmatrix}
\end{equation*}
is a symmetric matrix of bandwidths for all $p$ managements, as well as the $(p^2 - p) / 2$ correlations between managements.
For the multiple variance version (Figure \ref{fig:models}d, MB-MV-GK):
\begin{equation*}
	\mathbb{V}\text{ar}\left(\mathbf{u}\right) = \s\s^{\top} \circ \Big[\left(\C_M \otimes \mathbf{J}_q\right) \circ e^{-\mathbf{H}\otimes\D_E}\Big] \otimes \K_G \text{.}
\end{equation*}
Note that although we used $\mathbf{J}_q$ combined with Kronecker and Hadamard products above for clarity, these covariance structures can also be expressed using the Khatri-Rao product \citep{Liu_1999}, as shown in Figure \ref{fig:models}.



\subsection{Variance partitioning}\label{sec:LOF_variance}
Our initial interest was the separation of G$\times$E$\times$M interaction variances into components explained by (latent) environmental covariables and components attributed to diagonal lack of fit (LOF) effects.
To do so, we fitted a nested random effect with diagonal covariance structure on E$\times$M combinations in addition to the kernel-based random effects described in Section \ref{sec:models}.
These models are identifiable due to the structure imposed by the linear kernel $\K_E$ or distance matrix $\D_E$.
Summing these two independent random effects produces the following covariance model:
\begin{equation}\label{eq:LOF_variance}
	\mathbb{V}\text{ar}\left(\dot{\mathbf{u}}\right) = \mathbb{V}\text{ar}\left(\mathbf{u}\right) + \mathbb{V}\text{ar}\left(\mathbf{u}_{LOF}\right) = \boldsymbol{\Phi} + \mathbf{S} \otimes \K_G \text{,}
\end{equation}
where $\boldsymbol{\Phi}$ is one of the SV-LK, SV-GK, MV-LK, MV-GK, MB-SV-GK, or MB-MV-GK covariance structures described in Section \ref{sec:models} and $\mathbf{S}$ is a $pq\times pq$ diagonal covariance matrix containing lack of fit variances $\sigma^2_{lof|ij}$.
The Kronecker product $\mathbf{S} \otimes \K_G$ thus corresponds to the random LOF effect capturing the G$\times$E$\times$M variances not explained by the environmental covariables for each E$\times$M combination $l$.
For the single bandwidth models, the LOF term additionally captures variance not captured by the environmental kernel due to the separable structure for the E$\times$M interactions.
The separable structure for E$\times$M assumes the correlation between any pair of environments is equal for all managements.
While limiting in some sense, this assumption would allow us to separate environment and management even if they are fully confounded.
Note that the multi-bandwidth Gaussian kernel models relax this assumption by estimating multiple bandwidths, allowing for the G$\times$E structure to vary between managements.

We also fitted factor analytic models of orders $1$, $2$, and $3$ where we regarded the diagonal elements of $\LA\LA^\top$ as variances captured by latent environmental covariables and the diagonal elements of $\PS$ as LOF variances.

Finally, we fitted two baseline models.
Like all other models, the first baseline model contains a fixed effect for all E$\times$M combinations.
It also contains a random genetic main effect $\mathbf{u}_g \sim \mathcal{N}\left(\mathbf{0}, \sigma^2_g\K_G\right)$, two random two-way interactions $\mathbf{u}_{gm} \sim \mathcal{N}\left(\mathbf{0}, \sigma^2_{gm}\I_{p} \otimes \K_G\right)$ and $\mathbf{u}_{ge} \sim \mathcal{N}\left(\mathbf{0}, \sigma^2_{ge}\I_{q} \otimes \K_G\right)$, and the three-way interaction $\mathbf{u}_{gem} \sim \mathcal{N}\left(\mathbf{0}, \sigma^2_{gem}\I_{pq} \otimes \K_G\right)$.
This baseline model fully partitions the additive genetic variance in main effect, two-way interactions, and three-way interaction, and we thus refer to it as the full variance partitioning (FVP) model.
The second baseline model contains an additive genetic main effect as described in Section \ref{model:ME} and lack of fit term which is equivalent to the MDe model described by \citet{BandeiraeSousaetal_2017}.
Note that \citet{BandeiraeSousaetal_2017} use the term deviation where we use lack of fit.
Also note that while this model contains both a main effect and interaction term, no correlations between environments or managements are estimated.
In fact, the main effect model assumes correlations between all E$\times$M combinations are $1$ and the covariance matrix of the interaction term is a Kronecker product of a heterogeneous diagonal covariance structure for E$\times$M combinations and a kinship matrix, which is equivalent to $\mathbb{V}\text{ar}\left(\mathbf{u}_{LOF}\right)$ as described in (\ref{eq:LOF_variance}).

All reported variances were averaged over environments if applicable.
For example, the lack of fit variances reported for each management $i$ were obtained as:
\begin{equation*}
	\bar{\sigma}^2_{LOF|i} = \dfrac{1}{q}\sum_{j=1}^{q} s_{j|i} \text{,}
\end{equation*}
where $s_{j|i}$ is the $j$th element of that part of the diagonal of $\mathbf{S}$ that corresponds to management $i$.
Management-specific lack of fit variances for the FA models were obtained in a similar fashion by averaging the relevant diagonal elements of $\PS$.
Variances absorbed by the random effects based on (latent) covariables were obtained by averaging the diagonal elements of $\LA\LA^\top$ for the FA models and $\s\s^\top$ for the multiple-variance kernel models.
Variances associated with the two- and three-way interaction effects of the FVP model, genetic main effects of the FVP and ADD models, and kernel-based effects of the single-variance kernel models did not need to be averaged as these models did not contain environment-specific variances.

Similarly, all reported heritabilities were based on variances that were averaged over environments, if applicable, and computed as a ratio of additive genetic variance over phenotypic variance in lieu of the traditional measure defined by \citet[p.~53]{Walshetal_2025}.
They thus represent an average within-environment heritability, rather than heritabilities of means across environments:
\begin{equation}\label{eq:h2}
	\bar{h}^2_i = \dfrac{\bar{\sigma}^2_{a|i}}{\bar{\sigma}^2_{a|i} + \bar{\upsilon}_i} \text{,}
\end{equation}
where $\bar{\sigma}^2_{a|i} = \bar{\sigma}^2_{i} + \bar{\sigma}^2_{LOF|i}$ is the total additive genetic variance for management $i$ obtained by summing the mean lack of fit variance $\bar{\sigma}^2_{LOF|i}$ and mean variance obtained from $\s\s^\top$ or $\LA\LA^\top$ depending on the model ($\bar{\sigma}^2_{i}$).
Note that for the FVP model $\bar{\sigma}^2_{LOF|i}$ is simply the sum of the variance components associated with the two- and three-way interaction effects, and that $\bar{\sigma}^2_{i}$ is the variance component of the genetic main effect for the ADD and FVP models.
Finally, $\bar{\upsilon}_i$ is the mean variance of all BLUEs for management $i$ obtained from the first stage analyses.
Under the assumption of covariances between BLUEs being $0$, as implied by the diagonal weights approximation we used for the second stage analyses, (\ref{eq:h2}) is thus similar to the \textit{ad hoc} heritability measure based on effective error variances described by \citet{PiephoandMohring_2007}.

\subsection{Cross-validation setup for sparse testing genomic prediction}\label{sec:MM:GP}
We evaluated genomic prediction accuracy of the different models described in Section \ref{sec:models} in sparse testing scenarios with varying degrees of sparsity.
The goal was thus to predict the non-observed genotype by environment by management combinations that were missing from the phenotypic data.
\citet{Jarquinetal_2020} provide more details on sparse testing and the associated prediction setting.
Evaluating genomic prediction accuracy within this prediction scenario allows us to investigate the impact of modeling heterogeneous genetic variances as we have phenotypic data for every environment and management.
The fact that we have phenotypic data for all $E\times M$ combinations, and are thus not extrapolating to new environments, also allows for including their main effects as fixed.
Furthermore, the sparse testing aspect allows for evaluating the usefulness of environmental covariables relative to a factor analytic model as the connectivity between environments changes.

We note, however, that the models presented in this paper are by no means applicable to this prediction scenario alone.
Both the linear and Gaussian kernel-based models can be easily applied to scenarios where the goal is to make predictions for nested G$\times$E or G$\times$E$\times$M effects for new environments, new years, new locations, or any of the other settings discussed by, e.g., \citet{Hrachovetal_2026}.
To do so, environmental covariables, either observed for new locations or predicted based on, e.g., long-term means for new years, are used to define the correlations with environments in the TPE sample.
For linear kernels these correlations can be computed directly, while for Gaussian kernels they can be obtained using the bandwidth estimated from the training data.
The one caveat with extrapolation into new locations or years is that for the multiple-variance models, genetic variances for new environments cannot be estimated as there is no data to do so.
In these other prediction scenarios, environments may be taken as random or separated in locations and years as needed.
Additionally, predictions for new genotypes can be obtained in all scenarios as our kernel-based models still incorporate the standard kinship matrix.
We again note that the ability of the kernel-based models to extrapolate to new environments is an advantage over the FA models considered here as the factors for the latter are latent, and thus cannot be estimated for new environments.

We varied the level of sparsity within the sparse testing scenarios by sampling increasingly large groups of genotypes which were considered as checks, i.e., genotypes for which phenotypic data from all environments and managements were available.
The remaining genotypes had phenotypic data from only $2$ environments and, together with the checks, formed the training set.
The training sets were sampled such that the same number of training records was available for each E$\times$M combination.
The test set thus consisted of the combinations of genotypes, managements, and environments for which phenotypic data were missing.
Performance was evaluated for each E$\times$M combination in terms of correlation and root mean squared error (RMSE) between BLUPs from the LMMs and centered test set phenotypes \citep{Runcieandcheng_2019}.

\subsection{Software}\label{sec:MM:software}
All models were implemented using {\ttfamily ASReml-R} version {\ttfamily 4.2} \citep{butleretal_2023}.
We used either $1$, $2$, or $3$ factors for the FA models.
The kernel-based models were implemented using {\ttfamily ASReml-R}'s {\ttfamily own()} function which requires partial derivatives of the covariance matrices with respect to all parameters for the product of managements and environments.
Information on the use of {\ttfamily own()} as well as expressions for the partial derivatives can be found in the \hyperref[sec:appendix]{Appendix}.
Functions facilitating the implementation of the kernel-based models for G$\times$E and G$\times$E$\times$M data using {\ttfamily ASReml-R} are provided in the {\ttfamily R}-package {\ttfamily cornfruit} available at \url{https://github.com/KillianMelsen/cornfruit}.
Scripts and data to reproduce the presented results are available at \url{https://github.com/KillianMelsen/kernels_2026}.

\begin{figure}
	\begin{tikzpicture}
		\node[inner sep=0pt] (pic) at (0, 0) {\includegraphics[width=\textwidth]{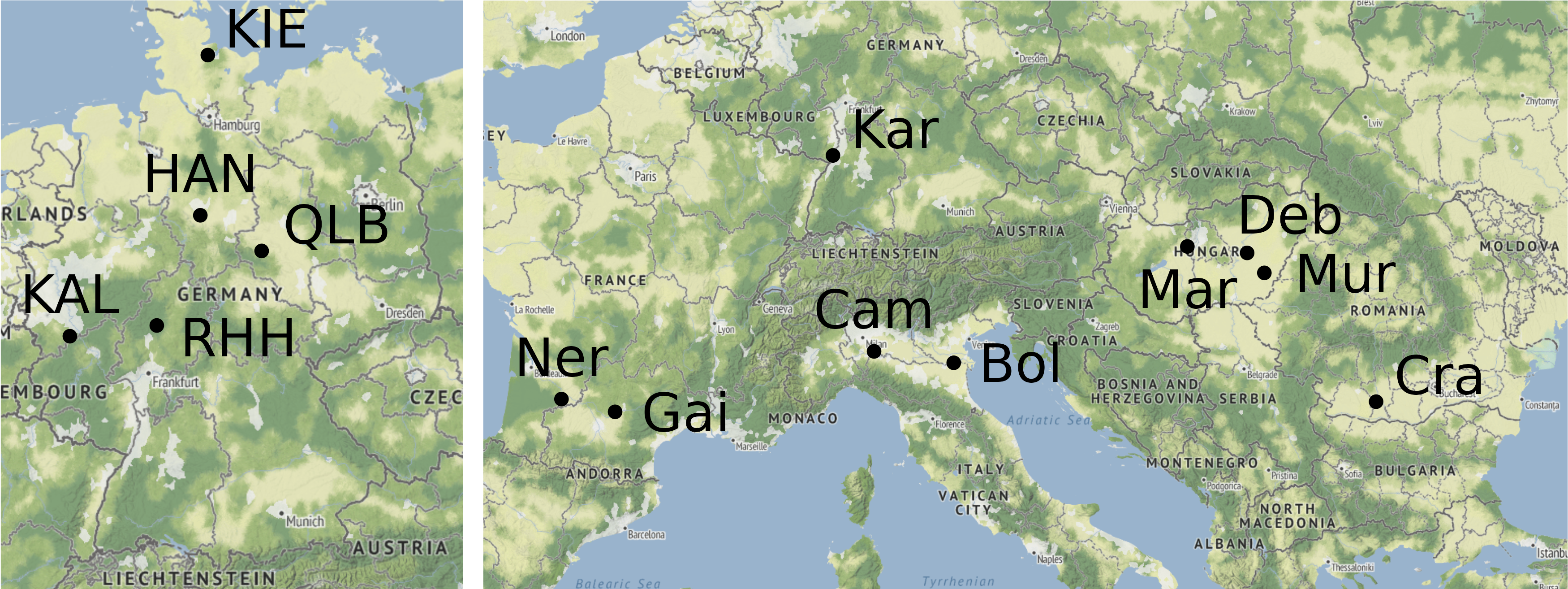}};
		\node (A) at (-7, 2.4) {\LARGE \textbf{a}};
		\node (B) at (-2.5, 2.4) {\LARGE \textbf{b}};
	\end{tikzpicture}
	\caption{The locations of field trials in the BRIWECS (\textbf{a}) and DROPS (\textbf{b}) datasets after subsetting to a balanced set of records. The DROPS location Graneros (Gra), Chile is not shown.
	\\[5pt]
	\textbf{Alt text:} A map of Germany and central Europe showing the locations of the trials for the BRIWECS and DROPS datasets.}
	\label{fig:locations}
\end{figure}

\subsection{BRIWECS dataset}\label{sec:MM:BRIWECS}
The first G$\times$E$\times$M dataset we used to illustrate prediction performance of the different models is the BRIWECS dataset (Breeding innovations in wheat for efficient cropping systems, \citep{Wangetal_2025}).
It contains measurements on grain yield, among other traits, for $6$ locations in Germany during 2015-2020 for $228$ genotypes (Figure \ref{fig:locations}a).
In each environment (location-year), up to $9$ management practices were evaluated in independent co-located trials.
We used the $135$K SNP markers published by \citet{Lichthardtetal_2019} to compute a genomic relationship matrix using the {\ttfamily statgenGWAS} {\ttfamily R}-package (identity by state method and minor allele frequency threshold of $0.05$) after recoding using the {\ttfamily ASRgenomics} {\ttfamily R}-package \citep{vanRossumandKruijer_2025, Gezanetal_2022}.
We randomly sampled $250$ training and test sets for each level of sparsity in the sparse testing genomic prediction scenario.

\subsubsection{Pre-processing of phenotypic data}
We used a subset of the full dataset for our analyses to ensure we had a fully balanced dataset that could be used for evaluating different degrees of sparsity.
We initially considered the years 2015-2017 and the LN\_NF\_RF, HN\_NF\_RF, LN\_WF\_RF, and HN\_WF\_RF managements.
In the abbreviations for the managements, LN/HN stand for low and high nitrogen, NF/WF for trials without and with fungicide applications, and RF for rain-fed.
We discarded the RHH\_2017, GGE\_2015, GGE\_2016, and GGE\_2017 environments.
RHH\_2017 had data for only $50$ genotypes and GGE\_2015, GGE\_2016, and GGE\_2017 had data for only a single management.

We fitted first stage, single-trial models to the remaining $14$ environments to obtain BLUEs and weights using the {\ttfamily R}-package {\ttfamily LMMsolver} \citep{Boer_2023}:
\begin{equation*}
	y_{ikuvw} = \mu + g_{ik} + b_{iu} + r_{v} + c_{w} + f\left(r, c\right)_{vw} + \epsilon_{ikuvw} \text{,}
\end{equation*}
where $g_{ik}$ is a nested fixed effect for each genotype $k$ within each management $i$, $b_{iu} \sim \mathcal{N}\left(0, \sigma_{b}^2\right)$ is a random effect for each management-block combination, and $r_{v} \sim \mathcal{N}\left(0, \sigma_{r}^2\right)$ and $c_{w} \sim \mathcal{N}\left(0, \sigma_{c}^2\right)$ are random effects for the rows and columns.
Spatial trends across rows and columns are modeled by the smooth surface $f\left(r,c\right)$ \citep{Rodriguez-Alvarezetal_2018}.
The overall intercept is given by $\mu$ and $\epsilon_{ikuvw} \sim \mathcal{N}\left(0, \sigma_{\epsilon}^2\right)$ is the residual.
See \citep{Wangetal_2025} for more details on the experimental design of individual trials.
BLUEs for each management-genotype combination for the second stage G$\times$E$\times$M analyses were then obtained as $\bar{y}_{ik} = \mu + g_{ik}$.
Weights were obtained as the diagonal elements of the precision matrix of BLUEs for each environment by management combination \citep{Damesaetal_2017, Smithetal_2001}.
We subset the BLUEs to those corresponding to the LN\_NF\_RF and HN\_NF\_RF managements, i.e., rain-fed trials with low and high nitrogen, no fungicide application, for a total of $28$ E$\times$M combinations.
We finally discarded any genotypes that were not evaluated in all $28$ E$\times$M combinations, resulting in a final balanced dataset of $5656$ BLUEs and associated weights for $202$ genotypes.

\subsubsection{Pre-processing of environmental data and kernels}
We used the NASA POWER database and {\ttfamily R}-package {\ttfamily EnvRtype} to obtain environmental covariables for all $14$ environments \citep{Costa-Netoetal_2021b}.
This requires geographical coordinates, as well as start and end dates for the period over which data will be retrieved.
We used the average sowing and harvest dates for each environment to define the periods.
Four environments had missing harvest dates.
We computed the average trial duration using the other $10$ environments and used this average combined with the sowing dates to obtain estimated harvest dates for the four environments.
We kept $18$ of the environmental covariables obtained from POWER, including covariables related to precipitation, solar radiation, and temperature.
See Table \ref{SM:BRIWECS_table} of the Supplementary Material (SM) for the geographical coordinates, sowing and harvest dates, and environmental covariables we used to compute the kernels.

We transformed the time-axis of the obtained covariables from days after sowing to thermal time ($\tau$, growing degree days) using the following formula \citep{NDAWN_2025a}:
\begin{equation}\label{eq:thermal_time}
	\tau_a = \dfrac{T_{min|a} + T_{max|a}}{2} - 32 \text{,}
\end{equation}
where the thermal time $\tau$ on day $a$ is computed using the minimum and maximum temperatures of that day and a base temperature of $32$ degrees Fahrenheit.
If $T_{min|a}$ or $T_{max|a}$ were below $32$ or above $95$, the values were set to the minimum and maximum temperatures of $32$ or $95$, respectively.
We then scaled and centered the environmental covariables before computing the mean value for every covariable for windows of $100$ accumulated growing degree days.
These means were then used as features to compute the kernels using (\ref{eq:linear_kernel}) and (\ref{eq:gaussian_kernel}).

\subsection{DROPS dataset}\label{sec:MM:DROPS}
The second dataset we use to illustrate the benchmark and kernel-based G$\times$E$\times$M models is the DROPS dataset (Drought-tolerant yielding plants, \citealt{milletetal_2019, milletetal_2016}).
It contains yield data on $246$ maize hybrids in several locations across Europe and one in Chile in 2012 and 2013 under both irrigated and rain-fed watering regimes (Figure \ref{fig:locations}b).
We computed the genomic relationship matrix using the $50$K SNP markers available in the DROPS data repository (\url{https://doi.org/10.15454/IASSTN}) and the {\ttfamily statgenGWAS} {\ttfamily R}-package, filtering any SNP markers with a minor allele frequency under $0.05$ and using the identity by state method \citep{vanRossumandKruijer_2025}.
We randomly sampled $150$ training and test sets for the sparse testing genomic prediction scenarios.

\subsubsection{Pre-processing of phenotypic data}
We again used a subset of the full dataset for our analyses to obtain a fully balanced dataset.
We discarded the Deb13 environment as it did not contain an irrigated trial.
We then fitted first stage models to the remaining $28$ environment by management combinations to obtain BLUEs and weights using {\ttfamily LMMsolver}:
\begin{equation*}
	y_{kuvw} = \mu + g_{ik} + b_{iu} + r_{v} + c_{w} + f\left(r, c\right)_{vw} + \epsilon_{ikuvw} \text{,}
\end{equation*}
where $g_{k}$ is a fixed effect for each genotype $k$ within a given E$\times$M combination, $b_{u} \sim \mathcal{N}\left(0, \sigma_{b}^2\right)$ is a random replicate effect, and $r_{v} \sim \mathcal{N}\left(0, \sigma_{r}^2\right)$ and $c_{w} \sim \mathcal{N}\left(0, \sigma_{c}^2\right)$ are random effects for the rows and columns.
Spatial trends across rows and columns are modeled by the smooth surface $f\left(r,c\right)$ \citep{Rodriguez-Alvarezetal_2018}.
The overall intercept is given by $\mu$ and $\epsilon_{kuvw} \sim \mathcal{N}\left(0, \sigma_{\epsilon}^2\right)$ is the residual.
We refer to the paper by \citet{milletetal_2016} for more details on the experimental design.
BLUEs were then obtained as $\bar{y}_{k} = \mu + g_{k}$ and weights were obtained as described for the BRIWECS data.
We discarded any genotypes that were not evaluated in all $28$ E$\times$M combinations, resulting in the final balanced dataset of $6552$ BLUEs and weights for $234$ genotypes.

\subsubsection{Pre-processing of environmental data and kernels}
The DROPS data repository contains daily measurements from sowing to harvest on environmental covariables obtained from weather stations and other sensors.
Like for the BRIWECS data, we first transformed the time-axis of these covariables from days to accumulated thermal time.
We used equation (\ref{eq:thermal_time}), but changed the minimum (base) and maximum temperatures to $50$ and $86$ degrees Fahrenheit \citep{NDAWN_2025b}.
We used $12$ of the environmental covariables to construct the kernels: RHmin.air, RHmax.air, RHmean.air, Raincum, Windspeedmax, ET0.air, Rad, Ri, VPD.air, VPD.apex, Tmax.apex, and Tnight.
See the DROPS data repository for details on these covariables.
We again scaled and centered the environmental covariables before computing the mean values for windows of $100$ growing degree days and used these means as features to construct the kernels using (\ref{eq:linear_kernel}) and (\ref{eq:gaussian_kernel}).

\section{Results and discussion}\label{sec:RD}
\subsection{BRIWECS results}
We initially fitted LMMs to investigate the partitioning of variance as described in Section \ref{sec:LOF_variance}.
On average, mean narrow-sense heritabilities considering the total additive genetic variance, i.e., the sum of genetic variances captured by the structured and LOF effects, were high at $0.90$ for the high nitrogen management and $0.84$ for the low nitrogen management.
Depending on the specific model used, values ranged between $0.88$ and $0.91$ for the high nitrogen management and $0.82$ - $0.88$ for the low nitrogen management (see Table \ref{SM:h2_table} of the SM).
While the original paper describing the data did not include any estimates for heritabilities, a publication on a similar G$\times$E dataset on historic wheat cultivars evaluated in Germany and northern France reported high heritabilities as well \citep{Gognaetal_2022}.
Total genetic variance was consistently higher in the high nitrogen management ($89.33\text{ dt}^2/\text{ha}^2$ averaged over the different models and environments) compared to the low nitrogen management ($54.76\text{ dt}^2/\text{ha}^2$).

\begin{figure}[h]
	\centering
	\includegraphics[width=\textwidth]{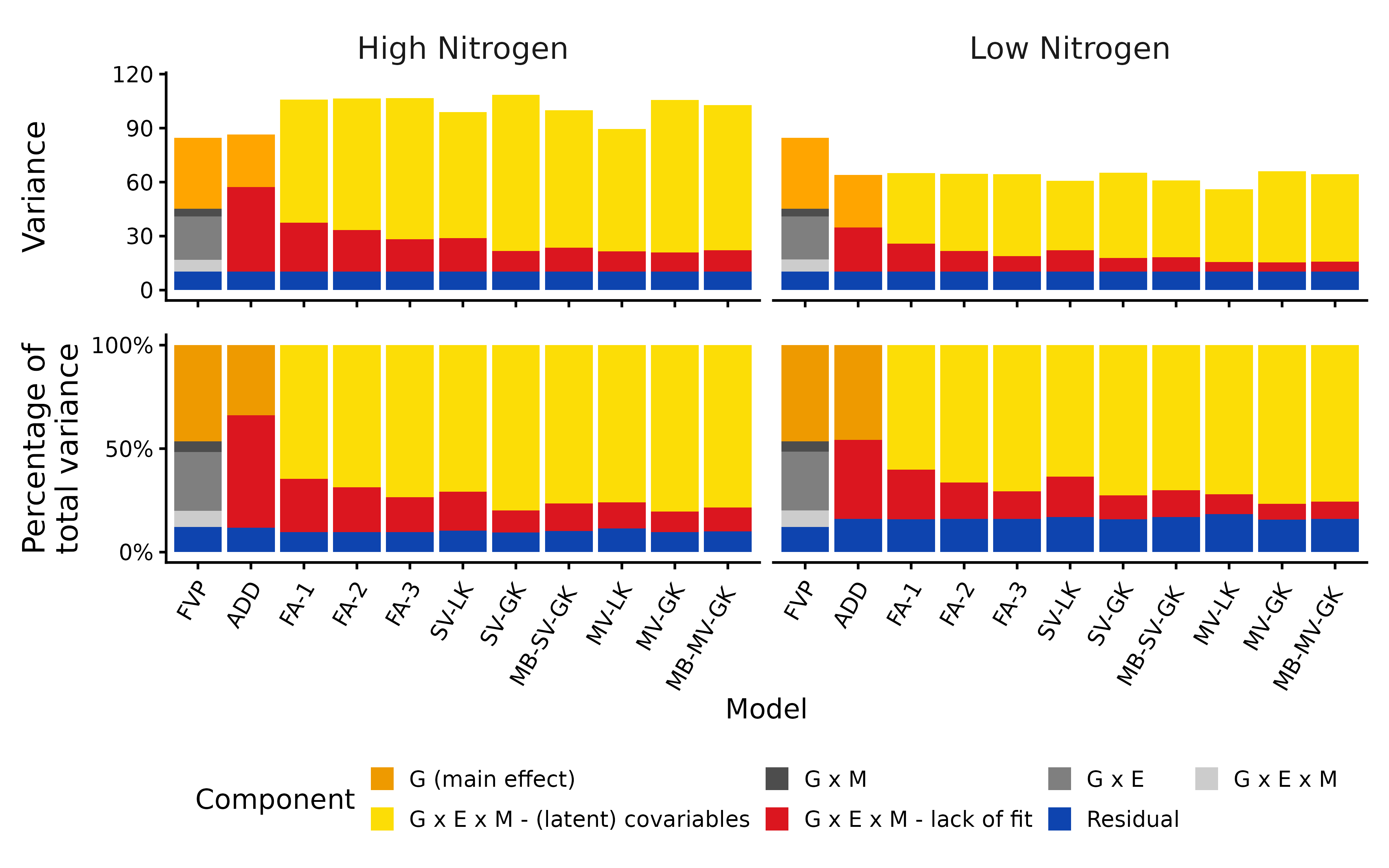}
	\caption{Partitioning of total variance of grain yield (overall means of $63.5$ and $58.1$ dt/ha for the high and low nitrogen managements, respectively) into variance captured by the genetic main effect, two- and three-way random interaction effects, nested G$\times$E$\times$M effects involving (latent) environmental covariables, nested G$\times$E$\times$M lack of fit effects, and residual effects, averaged over the $14$ environments and shown for each management of the BRIWECS dataset.
	Note that the yellow and red bars represent nested G$\times$E$\times$M effects while the gray bars represent two- and three-way interaction effects.
	\\[5pt]
	\textbf{Alt text:} Bar charts showing the amounts and percentages of variance explained by the different effects in the variance partitioning models for the BRIWECS dataset.}
	\label{fig:BRIWECS_LOF}
\end{figure}

While heritabilities were very similar between the different models, proportions of the total genetic variance absorbed by the (latent) covariables differed significantly (Figure \ref{fig:BRIWECS_LOF}).
Averaged over environments and managements, the covariables explained $78$\% of the genetic variance when using the single variance linear kernel, $87$\% when using the multiple variance linear kernel, $87$\% for the single variance Gaussian kernel, and $90$\% for the multiple variance Gaussian kernel.
The additional flexibility of the multiple bandwidth models did not result in a further increase.
In fact, the percentages for the MB-SV-GK and MB-MV-GK models were slightly lower at $85$\% and $88$\%, respectively.
There is a similar trend for the factor analytic models where the percentage of genetic variance explained by the latent environmental covariables increased from $71\%$ for the FA-1 model, to $77\%$ for the FA-2 model and finally $82\%$ for the FA-3 model.
For the baseline FVP and ADD models the percentage of genetic variance explained by the main effect was low at $53$\% and $45$\%, respectively.
Introducing the additional flexibility of the Gaussian kernel and environment-specific rates of shrinkage thus increased the explanatory power of the environmental covariables, but allowing for management-specific bandwidths did not cause a further increase.
Charts showing the variance partitioning for individual environments can be found in Figure \ref{fig:BRIWECS_LOF_EXTRA_A} and \ref{fig:BRIWECS_LOF_EXTRA_B} of the SM.

The full variance partitioning reveals that the majority of interaction variance could be attributed to G$\times$E interactions as the variance components for the G$\times$M and G$\times$E$\times$M effects were both small (Figure \ref{fig:BRIWECS_LOF}).
The small sizes of these variance components imply that the G$\times$E structure was relatively stable across the levels of management, meaning a fully separable structure with a single bandwidth may be appropriate.

\begin{figure}[h]
	\centering
	\includegraphics[width=\textwidth]{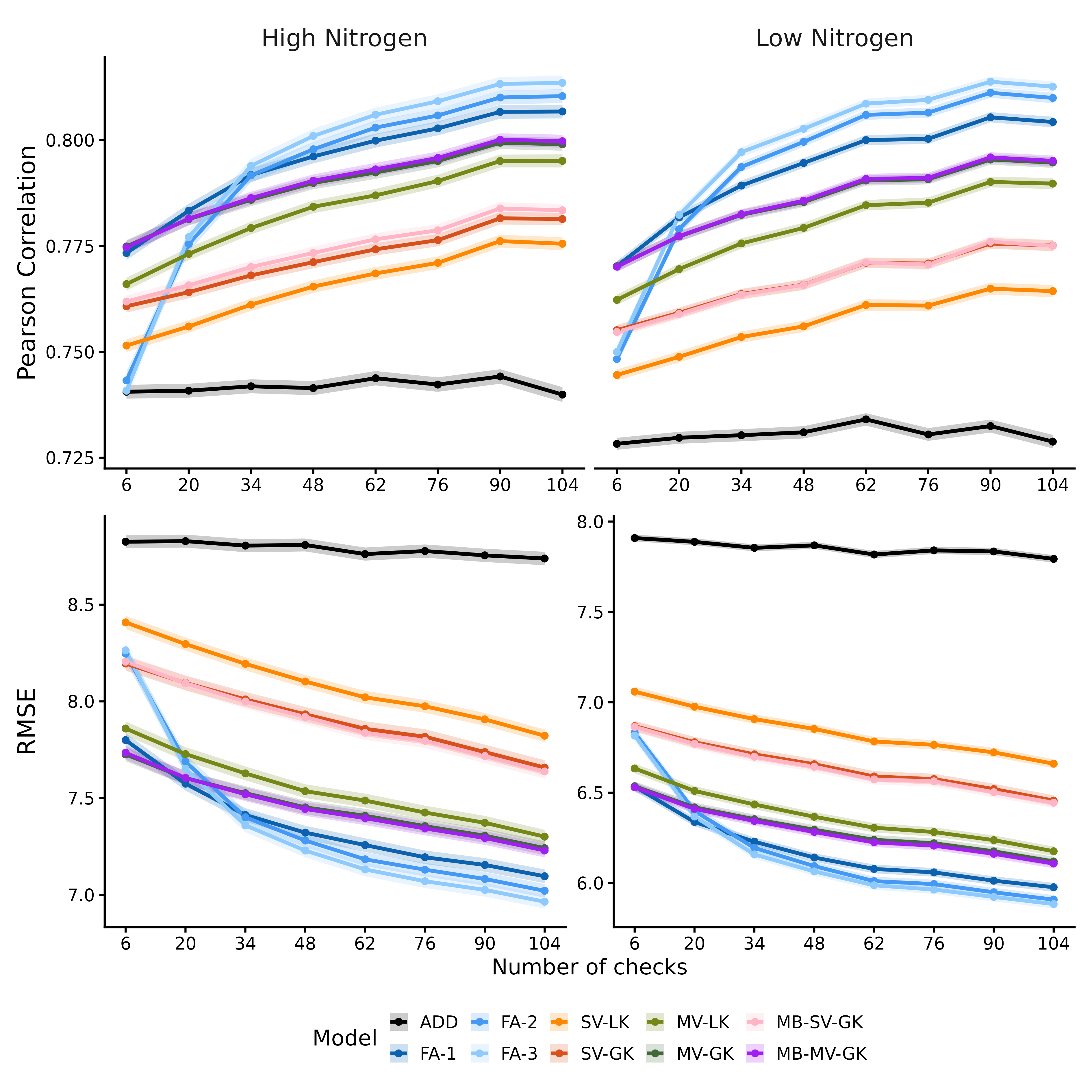}
	\caption{Pearson correlation and root mean squared error (RMSE) between BLUPs from the linear mixed models and centered test-set phenotypes for the BRIWECS dataset.
	Shaded areas represent standard errors.
	Accuracies are averaged over environments and shown separately for the low and high nitrogen managements.
	Checks are genotypes that are replicated in all environments.
	Note that as the number of checks increases, the sparsity in the sparse testing scenario decreases.
	Also note that the y-axes for RMSE have been scaled differently for clarity.
	Finally note that the lines for the MV-GK and MB-MV-GK models almost perfectly overlap.
	Enlarged versions of the individual plot panels are provided in SM Figures \ref{fig:BRIWECS_CV2_PCHN}--\ref{fig:BRIWECS_CV2_RMSELN}.
	\\[5pt]
	\textbf{Alt text:} Line graphs showing the genomic prediction accuracies obtained using the different models and levels of sparsity for the BRIWECS dataset.}
	\label{fig:BRIWECS_CV2}
\end{figure}

Next, we evaluated model performance in sparse testing genomic prediction scenarios with $8$ levels of sparsity.
Overall, the models incorporating G$\times$E$\times$M interactions outperformed the ADD model in terms of both correlation and RMSE (Figure \ref{fig:BRIWECS_CV2}).
Furthermore, there is a clear trend for the G$\times$E$\times$M models where performance increased as the level of sparsity decreased, with an improvement over the ADD model of about $8$\% in terms of correlation and $18.5$\% in terms of RMSE at the least sparse end.
The FA models of order $2$ and $3$ were most clearly affected by the lack of connectivity in the most sparse settings, where the performance dropped significantly.
The kernel-based models generally performed reasonably well, although there was a large difference between the single and multiple variance versions.
The multiple variance kernel models consistently outperformed the single variance versions by a significant margin.
The multiple variance Gaussian kernel models with single and multiple bandwidths seemed to perform similarly to the FA-1 model up to $20$ checks, although differences are small.
The higher order FA models clearly performed better after $34$ checks, with the difference compared to the MV-GK and MB-MV-GK models increasing as the level of sparsity decreased.
The addition of management-specific bandwidths in the MB-SV-GK and MB-MV-GK models did not result in any increase in prediction performance.
This agrees with the variance partitioning results that implied stable G$\times$E across managements, meaning there is no need for multiple bandwidths.
Graphs showing genomic prediction accuracies for the different models and levels of sparsity for individual environments are provided in SM Figures \ref{fig:BRIWECS_CORR_EXTRA_A}--\ref{fig:BRIWECS_RMSE_EXTRA_B}.

\subsection{DROPS results}
\begin{figure}[h]
	\centering
	\includegraphics[width=\textwidth]{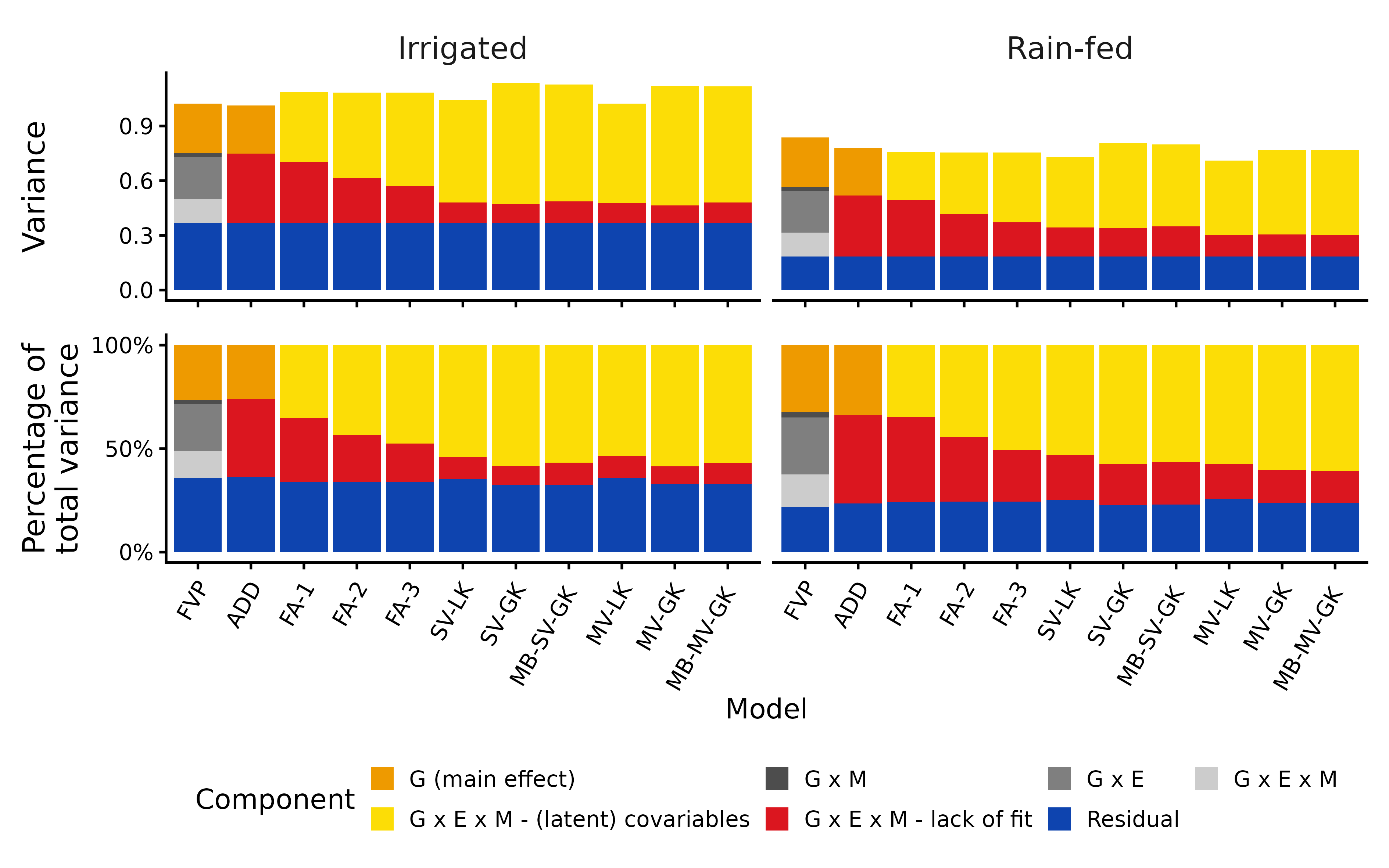}
	\caption{Partitioning of total variance of grain yield (overall means of $5.37$ and $7.62$ t/ha for the rain-fed and irrigated managements, respectively) into variance captured by the genetic main effect, two- and three-way random interaction effects, nested G$\times$E$\times$M effects involving (latent) environmental covariables, nested G$\times$E$\times$M lack of fit effects, and residual effects, averaged over the $14$ environments and shown for each management of the DROPS dataset.
	Note that the yellow and red bars represent nested G$\times$E$\times$M effects while the gray bars represent two- and three-way interaction effects.
	\\[5pt]
	\textbf{Alt text:} Bar charts showing the amounts and percentages of variance explained by the different effects in the variance partitioning models for the DROPS dataset.}
	\label{fig:DROPS_LOF}
\end{figure}

Like for the BRIWECS data, we fitted LMMs including an additional random effect with diagonal covariance structure as described in Section \ref{sec:LOF_variance} to investigate the variance partitioning of the different models.
Mean heritabilities for the DROPS dataset were significantly lower than for BRIWECS at around $0.66$ for the irrigated management and $0.76$ for the rain-fed management.
Depending on the model used the values ranged between $0.64$ - $0.68$ for the irrigated management and $0.74$ - $0.78$ for the rain-fed management (see SM Table \ref{SM:h2_table}).
Mean total genetic variance was larger for the irrigated management ($0.709\text{ t}^2/\text{ha}^2$) than for the rain-fed management ($0.586\text{ t}^2/\text{ha}^2$, Figure \ref{fig:DROPS_LOF}).

While estimated heritabilities were again similar between models, the percentages of genetic variance explained by the covariables followed a trend similar to what was found in the BRIWECS dataset.
For the factor analytic models the percentages increased from $50\%$ for the FA-1 model, to $63\%$ for the FA-2 and $70\%$ for the FA-3 model.
Similarly, the explanatory power of the environmental covariables increased as the complexity of the kernels increased.
Percentages of genetic variance explained increased from $78\%$ for the SV-LK model, to $81\%$ for the SV-GK and MV-LK models, and $84$\% for the MV-GK model.
Percentages were similar for the multiple bandwidth models at $79$\% for the MB-SV-GK model and $83$\% for the MB-MV-GK model.
For the baseline FVP and ADD models the percentages of genetic variance explained by the main effect were again low at $41$\% and $42$\%, respectively.

The full variance partitioning model again shows that G$\times$M interactions were negligible.
While G$\times$E interactions were still the most important, the variance component associated with G$\times$E$\times$M interactions was relatively large as well.
This indicates that for the DROPS dataset, G$\times$E interactions changed across managements, meaning that the fully separable structure of the single bandwidth models may not be ideal.
Charts showing the variance partitioning for individual environments can be found in Figures \ref{fig:DROPS_LOF_EXTRA_A} and \ref{fig:DROPS_LOF_EXTRA_B} of the SM.

\begin{figure}[h]
	\centering
	\includegraphics[width=\textwidth]{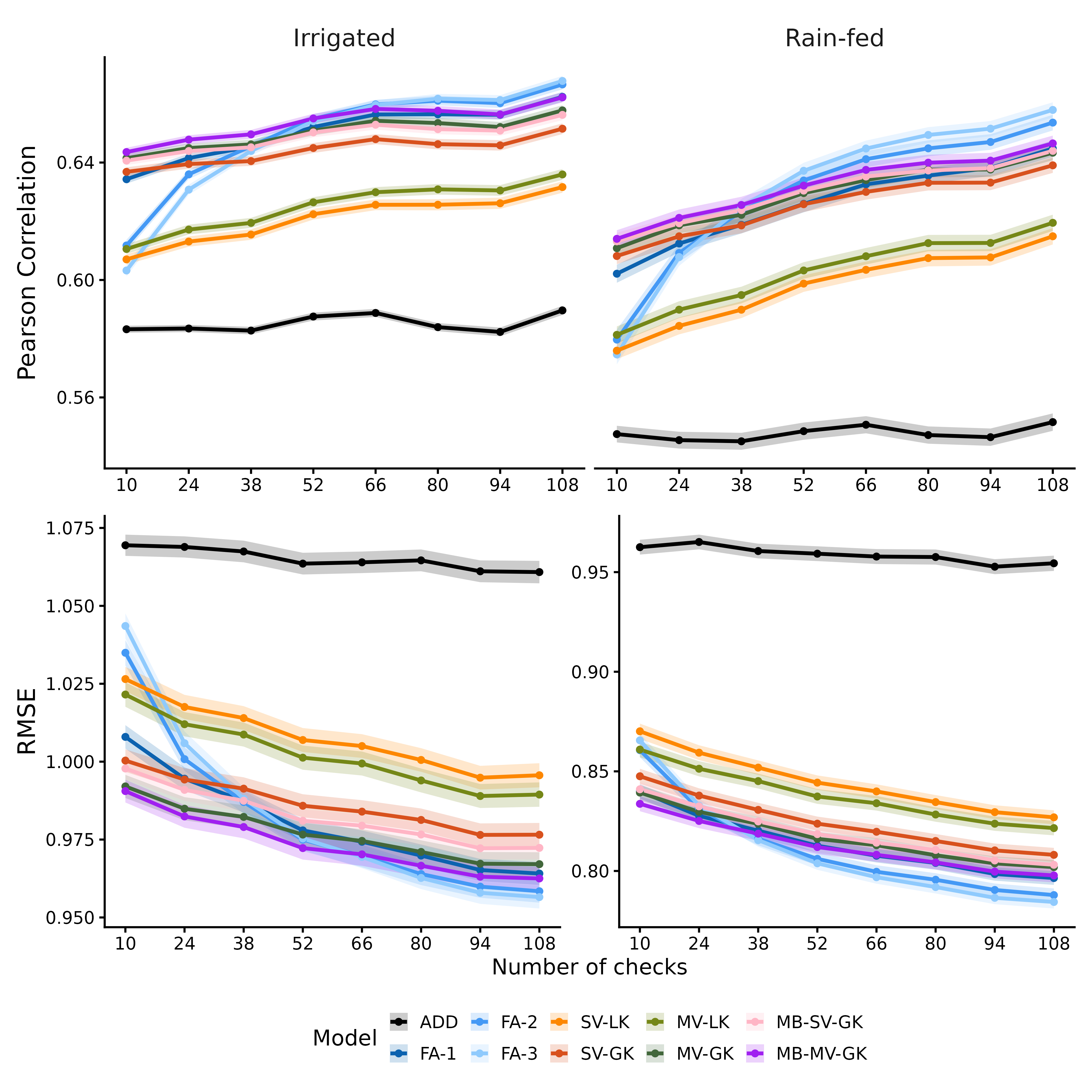}
	\caption{Pearson correlation and root mean squared error (RMSE) between BLUPs from the linear mixed models and centered test-set phenotypes for the DROPS dataset.
		Accuracies are averaged over environments and shown separately for the irrigated and rain-fed managements.
		Checks are genotypes that are replicated in all environments.
		Note that as the number of checks increases, the sparsity in the sparse testing scenario decreases.
		Also note that the y-axes for RMSE have been scaled differently for clarity.
		Enlarged versions of the individual plot panels are provided in SM Figures \ref{fig:DROPS_CV2_PCW}--\ref{fig:DROPS_CV2_RMSER}.
		\\[5pt]
		\textbf{Alt text:} Line graphs showing the genomic prediction accuracies obtained using the different models and levels of sparsity for the DROPS dataset.}
	\label{fig:DROPS_CV2}
\end{figure}

Results from the genomic prediction scenarios for the DROPS dataset were generally similar to those for the BRIWECS data.
The ADD model performed worse than all interaction models (Figure \ref{fig:DROPS_CV2}).
The FA-2 and FA-3 models again showed a significant drop in accuracy when little connectivity existed between environments.
The Gaussian kernels clearly outperformed the linear kernels in terms of correlation as well as RMSE, with the multiple variance Gaussian kernels performing slightly better than the single variance versions.
While the estimation of management-specific bandwidths did not seem to increase prediction performance for the BRIWECS data in terms of either correlation or RMSE, there was a slight increase in performance for the DROPS data with the MB-MV-GK model consistently being the best kernel model.
While the higher order FA models performed best when the number of checks is large, the MB-MV-GK model was better or equal up to between $38$ and $66$ checks depending on the management.

The increased performance of the multiple bandwidth models over the single bandwidth models corresponds well to the relatively large size of the G$\times$E$\times$M variance component in the FVP model, meaning that the G$\times$E structured depended on the management.
Indeed, on average the bandwidths for rain-fed and irrigated trials were $0.48$ and $0.37$, respectively, for the MB-MV-GK model.
Note that a lower bandwidth results in correlations closer to $1$, implying that G$\times$E interactions were stronger for the rain-fed trials than for the irrigated trials.
Finally note that the MB-SV-GK model failed to fit for two out of $150$ training sets at $10$ checks.
The MB-MV-GK model failed to fit for two training sets at $10$ checks and a single training set at $24$ checks.
These issues may have been caused by inappropriate starting values.
Graphs showing genomic prediction accuracies for the different models and levels of sparsity for individual environments are provided in SM Figures \ref{fig:DROPS_CORR_EXTRA_A}--\ref{fig:DROPS_RMSE_EXTRA_B}.

\subsection{Discussion}
G$\times$E$\times$M interactions are common in the breeding of field crops \citep{CooperandMessina_2023, Cooperetal_2022, deLeonetal_2016}.
Several statistical and machine learning models including these interactions in genomic predictions have been proposed, with LMMs involving REML estimation of structured covariance matrices and Bayesian implementations of kernel-based reaction norm models being two commonly used frameworks \citep{jarquinetal_2014, gilmouretal_1995}.
Here, we made an effort to combine aspects from both frameworks by introducing REML implementations of several kernel-based models with homo- and heterogeneous genetic variances for multi-environment, multi-management trial data.
We have shown that a nonlinear, multi-bandwidth Gaussian kernel combined with an unstructured model for managements and environment-specific variances improves G$\times$E$\times$M predictions compared to factor analytic or linear kernel models in sparse testing scenarios.
We have also shown that more complex kernel models generally increase the percentage of interaction variance explained by environmental covariables and that management-specific bandwidths allow for appropriate modeling of management-dependent G$\times$E.

Current implementations of kernel-based models are limited in the sense that they only allow for a single genetic variance across all environments for any level of management.
The benefit of modeling environment-specific genetic variances can be readily explained from a shrinkage and weighting of information perspective.
Homogeneous genetic variances across environments for a given management results in equal shrinkage if the residual variances are assumed to be equal.
Equal shrinkage implies equal weighting of information from different environments in the estimation of random effects.
This is clearly inappropriate if heterogeneity of genetic variance is present, and may result in lower predictive performance through increased bias.
Heterogeneity of genetic variance is well known to be a contributor to G$\times$E interactions \citep{vanEeuwijketal_2016}, and modeling environment-specific rates of shrinkage should thus improve predictive performance, as also shown by our results \citep{Okekeetal_2017}.

In addition to implementing kernel-based models with heterogeneous genetic variances, we provided REML implementations of single and multiple bandwidth Gaussian kernel-based models where we treat the bandwidths of the kernels as parameters to be estimated.
There have been several other approaches to choosing bandwidth values.
\citet{deloscamposetal_2010} proposed kernel averaging which involves including multiple random effects with different bandwidths.
The variance components associated with these random effects then act as weights, resulting in an overall kernel that is essentially a weighed average of the kernels associated with each effect.
A limitation of this approach is that the optimal bandwidth is approximated by summing multiple random effects, rather than estimated directly.
The addition of these extra random effects also increases the size of the mixed model equations, slowing down model fitting significantly.
Other methods used in the context of LMMs and closely related machine learning approaches like LS-SVM and Gaussian processes typically involve computationally expensive cross-validation, some sort of grid-search, optimization before fitting the model, or are restricted to simple GBLUP models (\citealp[Section 2.4]{Kocijan_2016}; \citealp{Perez-Elizaldeetal_2015, Endelman_2011, GianolaandvanKaam_2008}; \citealp[p.~75]{Suykensetal_2002}).
Finally we note that while bandwidths can be directly estimated using REML in the context of spatial models, software to do so is limited to the case of a single Gaussian kernel and no Kronecker products with kinship or other covariance matrices \citep{Piepho_2009}.
The REML estimation framework we proposed here extends this set of approaches and comes with the benefit that only a single random effect and no cross-validation or optimization prior to fitting the model is required.
Furthermore, REML estimation can be performed for models including complex interaction effects involving multiple management levels and environments at practically no additional computational cost.

While our implementation allows for combining the environmental kernels with different covariance structures for factors such as managements, the structure imposed by the environmental kernels also allows for fitting additional random effects for the environment.
Because of the structured and fixed nature of the linear kernels or distance matrices, the covariance models associated with these random effects can be unstructured.
In the current paper, we fitted an additional random lack of fit effect for the environments with a diagonal covariance structure to investigate what proportion of G$\times$E$\times$M variance was explained by the environmental covariables.
The results revealed that total genetic variance was higher under the favorable management conditions, i.e., high nitrogen and irrigated trials, which matches what is commonly found in plant breeding experiments \citep{Musimwaetal_2022, Beyeneetal_2016, Ud-Dinetal_1992}.
More interestingly, the additional random effect revealed that the environmental covariables explained a larger part of the G$\times$E$\times$M variance with increasing complexity of the kernel model.
In other words, Gaussian kernel models with heterogeneous genetic variances result in a larger fraction of interaction variance that is predictable given the environmental covariables.
Furthermore, the differences between bandwidth values for irrigated and rain-fed trials obtained using the multiple bandwidth Gaussian kernel models indicate that a fully separable structure may not be ideal in the presence of G$\times$E and G$\times$E$\times$M interactions.
Indeed, allowing for management-specific bandwidths increased prediction accuracy for the DROPS dataset where the full variance partitioning model indicated significant G$\times$E and G$\times$E$\times$M while G$\times$M was negligible.
For the BRIWECS data on the other hand, where G$\times$E$\times$M interactions were smaller, little benefit in estimating multiple bandwidths was observed.

While we only fitted an extra random effect with diagonal covariance matrix in the current paper, the structure in the kernels allows for extensions with random effects with more complex covariance matrices as well.
Our results with regards to variance partitioning show that both the kernel-based and FA models captured a significant part of the variance through their respective structures based on (latent) covariables.
If variance captured by the latent covariables is at least partly different from the variance captured by the observed covariables underlying the kernels, combining the FA and kernel structures would be an interesting way forward.
In fact, combining the kernels with a random effect with factor analytic covariance structure would lead to a different parameterization of the integrated factor analytic linear mixed model (IFA-LMM) proposed by \citet{Tolhurstetal_2022}.
Such models would offer a relatively straightforward implementation of what is essentially linear or nonlinear regression on known covariables, combined with linear regression on an additional set of latent environmental covariables.
Interaction variance and corresponding nested G$\times$E or G$\times$E$\times$M effects would then be naturally partitioned in (1) components predictable from environmental covariables, (2) components that are only observable in the current sample of the TPE, and (3) lack of fit components corresponding to the unique variances in $\PS$.
These last components can be thought of as environment-specific deviations that are neither predictable, nor explained by latent environmental covariables.
The second components associated with latent environmental covariables can be used to obtain predictions for seen environments, for example in sparse testing scenarios or for seen environments but unseen genotypes.
Finally, the first components can be used to extrapolate to new locations, new years, or combinations of new locations and years, potentially using long-term averages or quantiles of the environmental covariables \citep{Hrachovetal_2026}.
This would be especially interesting when combined with environmental covariables obtained from satellites, like we did for the BRIWECS data, as it would in principle allow for predictions to be made for any location on the planet.

Alternatively, the models discussed in this paper could be extended by including multiple kernels or kernels constructed using features retained after feature selection.
Currently, we constructed environmental kernels using features representing all combinations of environmental covariables and days.
We did not perform any feature selection, but it is well known that not all covariables and timeframes in the growth cycle are relevant \citep{milletetal_2019}.
Features representing different classes of covariables, e.g., temperature, radiation, or water availability could be used to construct different kernels.
Variances associated with each kernel would then indicate the relevance of the different sets of features for predicting G$\times$E \citep{deloscamposetal_2010}.
Similarly, kernels representing different growth stages could be included, or a kernel for environments could be combined with a nonlinear kernel for genotypes instead of a standard kinship matrix.

Finally, we would like to make some notes on our choice of modeling management as an additional categorical factor in the LMM.
There is some discussion on whether management should be regarded as a separate factor, or part of defining the environment.
\citet{Cooperetal_2022} reviewed the issue and provided a comprehensive overview of the different perspectives on modeling G$\times$E$\times$M data.
We would argue that when management is confounded with the environment, as in the case where each environment consists of a single trial run under standard local management practices (see, e.g., \citealp{Leeetal_2026, Ardayfioetal_2025}), management practices should indeed be viewed as defining the environment itself.
Within the context of kernels, this could be achieved by including management-specific covariables like nitrogen rate, or planting density in the matrix of environmental covariables used to construct the kernels \citep{Costa-Netoetal_2021a}.
One drawback of this G$\times$(E$\times$M) approach is that in the case of multiple differentially managed trials being conducted simultaneously at the same location, kernel-based correlations between these environment-management combinations will almost always be close to $1$.
This is a simple result of the fact that any meteorological covariables obtained from weather stations or satellites will be identical for these E$\times$M combinations.
We therefore believe that modeling management separately from environments is preferable if managements are included as an additional experimental factor crossed with environments, like in the datasets used in this paper.

Another benefit of modeling managements as a separate categorical factor is that any methodology developed for this purpose can be applied to multi-trait G$\times$E$\times$T data as well.
Joint G$\times$E analyses of traits like grain yield and protein content would be straightforward.
These models would also allow for integration of secondary traits derived from high-throughput phenotyping (HTP).
While modeling large numbers of HTP features directly is typically infeasible, a smaller number of features obtained after some form of dimension reduction can be modeled relatively easily.
Many dimension reduction methods exist, but approaches that produce independent features such as factor or principal component analysis seem most promising due to the lower number of correlations that need to be estimated and the ability of these secondary features to improve predictions for a focal trait of interest \citep{Melsenetal_2025}.

\section{Conclusion}\label{sec:C}
This paper provides easy-to-use REML implementations of linear and nonlinear kernel-based genomic prediction models for G$\times$E$\times$M data in standard mixed model software.
We have shown that extending the basic kernel models that, given a particular management, assume a single genetic variance for all environments to models with environment-specific variances results in higher prediction accuracies.
Furthermore, management-specific Gaussian kernels allow for more appropriate modeling of complex G$\times$E$\times$M interactions than the fully separable models with a single bandwidth.
We have also shown that Gaussian kernel models with heterogeneous variances capture more of the interaction variance than the single variance and linear versions, implying a larger proportion of G$\times$E$\times$M is predictable given the environmental covariables.
Finally, we have discussed some possible extensions to the different kernel-based models.
Feature selection or kernels representing different subsets of environmental features deserve attention, as the inclusion of non-relevant features in kernels may lead to sub-optimal predictions.
Furthermore, combining kernels with a factor analytic structure leads to models similar to the IFA-LMM model proposed by \citet{Tolhurstetal_2022}, separating the G$\times$E into predictable and observable components.
Finally, we also recognize that the modeling of multiple traits or HTP features in a G$\times$E context is effortlessly implemented in the framework discussed in this paper by estimating correlations between traits instead of managements, offering a relatively simple way of integrating phenomics, enviromics, and genomics in a single LMM.

\section*{Data availability}
Functions facilitating the implementation of the kernel-based models for G$\times$E and G$\times$E$\times$M data using {\ttfamily ASReml-R} are provided in the {\ttfamily R}-package {\ttfamily cornfruit} available at \url{https://github.com/KillianMelsen/cornfruit}.
All scripts and data required to reproduce the presented results are available at \url{https://github.com/KillianMelsen/kernels_2026}.
For those interested, additional details on the DROPS dataset as well as the raw data itself can be found in the repository (\url{https://doi.org/10.15454/IASSTN}) and accompanying publications \citep{milletetal_2019, milletetal_2016}.
Details and raw phenotypic data for the BRIWECS dataset can be found in the repository at \url{https://doi.org/10.6084/m9.figshare.27910269} and the publication by \cite{Wangetal_2025}.
SNP marker data is available in the Supplementary Material of \citet{Lichthardtetal_2019}.

\phantomsection
\label{sec:appendix}
\section*{Appendix: Programming framework and partial derivatives}
This Appendix contains a graphical overview and some information on the programming framework used to fit the kernel-based LMMs using the {\ttfamily own()} function of the {\ttfamily ASReml-R} software.
It also provides expressions for the partial derivatives of the different kernel-based covariance models required to fit them using {\ttfamily ASReml-R}.

\subsection*{Programming framework}
We make use of the {\ttfamily own()} function available in the {\ttfamily R}-package {\ttfamily asreml} ({\ttfamily ASReml-R} version {\ttfamily 4.2}) to fit the kernel-based LMMs.
The {\ttfamily own()} function allows specification of custom covariance models associated with random effects in the LMM through an external variance function.
It thus allows us to model the management and environment factors using either a single variance for each level of management, resulting in the same variance for all combinations of environments and that specific management, or unique variances for all environment-management combinations.
Furthermore, it allows us to model Gaussian kernels for which the bandwidth parameters are estimated directly using REML within complex models involving interactions.

The use of {\ttfamily own()} is illustrated in Figure \ref{fig:framework}.
Within the {\ttfamily asreml()} call, {\ttfamily own()} is used to specify the random part of the model, taking five arguments.
The argument {\ttfamily obj} specifies which factor in the data passed to {\ttfamily asreml()} should be modeled using the custom covariance structure, similarly to how the factor is specified for functions like {\ttfamily corgh()}, {\ttfamily fa()}, or {\ttfamily vm()}.
The {\ttfamily fun} argument takes a character variable referring to the name of the user-defined external variance function, e.g., {\ttfamily "varFun"}.
Finally, the {\ttfamily init}, {\ttfamily type}, and {\ttfamily con} arguments take vectors that allow specification of the initial values, types, and constraints for the parameters used to construct the custom covariance matrix.
We refer to the {\ttfamily ASReml-R} Reference Manual Version {\ttfamily 4.2} by \citet{butleretal_2023} for more information.

The external variance function itself must be defined in the {\ttfamily R} environment prior to calling {\ttfamily asreml()}.
It must have the arguments {\ttfamily order} and {\ttfamily kappa}, and no others.
While fitting the LMM, {\ttfamily asreml()} calls {\ttfamily own()} at every iteration, passing the number of levels of the factor specified using {\ttfamily obj} as {\ttfamily order} and current parameter estimates as {\ttfamily kappa}.
Note that for the first iteration, {\ttfamily kappa} will hold the initial values specified using {\ttfamily init}.
Inside the external variance function, {\ttfamily order} and {\ttfamily kappa} are then used to construct the covariance matrix and partial derivative matrices with respect to all parameters.
These are then returned as a list of matrices with the covariance matrix coming first, followed by all derivatives with respect to the parameters in the same order as {\ttfamily kappa}.
In other words, if the first element of {\ttfamily kappa} is the first variance, the second element in the returned list must be the derivative of the covariance matrix with respect to the first variance.

The {\ttfamily asreml()} function then uses the returned covariance matrix and derivatives to update the parameter estimates, before calling {\ttfamily own()} again.
The cycle illustrated in Figure \ref{fig:framework} continues in this way until convergence or the maximum number of iterations is reached.
The {\ttfamily own()} function provides a powerful interface to {\ttfamily ASReml-R}, allowing specification of any valid covariance structure for which derivatives can be obtained.
We provide expressions for the partial derivative matrices of the kernel-based covariance models in the remainder of this Appendix.

\begin{figure}[h]
	\centering
	\begin{tikzpicture}
			\node[rectangle, draw=black, very thick, minimum height=20, rounded corners] (VF) at (0, -0.4)
			{\ttfamily varFun(order, kappa)};
			\node[rectangle, draw=black, very thick, minimum height=20, rounded corners] (AS) at (0, 2)
			{\ttfamily asreml(random = $\sim$ own(obj, fun, init, type, con))};
			
			\draw[>= latex, ->, very thick] (-4, 1.65) to [out=-90,in=180] (VF);
			\draw[>= latex, ->, very thick] (VF) to [out=0,in=-90] (4, 1.65);
	
			\node[rectangle, align=center, draw=black, fill=white, very thick, rounded corners] at (3.27, 0.75)
			{Covariance matrix and\\partial derivatives};
			\node[rectangle, align=center, draw=black, fill=white, very thick, rounded corners] at (-4.16, 0.75)
			{Parameters};
		\end{tikzpicture}
	\caption{Programming framework for specifying custom covariance models in {\ttfamily ASReml-R}. The {\ttfamily own()} function can be used inside the {\ttfamily asreml()} call.. Arguments {\ttfamily obj}, {\ttfamily fun}, {\ttfamily init}, {\ttfamily type}, and {\ttfamily con} are used to specify the data factor, name of the custom external variance function, initial values for the (variance) parameters, types of parameters, and parameter constraints. The custom variance function itself must only have arguments {\ttfamily order} and {\ttfamily kappa} which take the number of factor levels and parameter estimates at a particular iteration, respectively, as passed by {\ttfamily ASReml-R} during model fitting.
	\\[5pt]
	\textbf{Alt text:} A graphical overview of how ASReml-R is used to fit the kernel-based linear mixed models.}
	\label{fig:framework}
\end{figure}

\subsection*{Partial derivatives}
This Section provides expressions for the partial derivatives of the different kernel-based covariance models with respect to the parameters used for the management and environment factors.
Note that the kinship matrix $\K_G$ has been left out as it is a constant and the same for all models.

\subsubsection*{Single variance linear kernel}
The parameters for the single variance linear kernel (SV-LK) model consist of $p$ genetic variances and $up/2$ genetic correlations where $u=p-1$:
\begin{equation*}
	\boldsymbol{\kappa} = \left[\sigma_1^2, \sigma_2^2, \dots, \sigma_p^2, \rho_{12}, \rho_{13}, \dots, \rho_{up}\right] \text{,}
\end{equation*}
resulting in the following derivatives with respect to the variances:
\begin{equation}\label{eq:d_v_svlk}
	\dfrac{\partial\left(\Covmat_M\otimes\K_E\right)}{\partial \sigma_i^2} = \dfrac{\partial \Covmat_M}{\partial \sigma_i^2}\otimes\K_E = \dfrac{\Covmat_M \circ \left(\mathbf{A}_i + \mathbf{A}_i^\top\right)}{\sigma_i^2}  \otimes \K_E \text{,}
\end{equation}
where $\mathbf{A}_i$ is a $p\times p$ matrix containing $0.5$ in the $i$th row, and zero everywhere else.
The SV-LK model produces the following derivatives with respect to the correlations:
\begin{equation}\label{eq:d_r_svlk}
	\dfrac{\partial\left(\Covmat_M\otimes\K_E\right)}{\partial \rho_{ii'}} = \dfrac{\partial \Covmat_M}{\partial \rho_{ii'}}\otimes\K_E = \left(\mathbf{s}_M\mathbf{s}_M^\top \circ \mathbf{B}_{ii'}\right) \otimes \K_E \text{,}
\end{equation}
where $\mathbf{B}_{ii'}\in\mathbb{R}^{p\times p}$ is an indicator matrix where $\mathbf{B}_{ii'} = \mathbf{B}_{i'i} = 1$ and all other elements are zero.
Note that for correlations $i \neq i'$.
The vector $\mathbf{s}_M$ contains standard deviations for each level of management so that $\mathbf{s}_M\mathbf{s}_M^\top \circ \Cormat_M = \Covmat_M$.

\subsubsection*{Multiple variance linear kernel}
The multiple variance linear kernel (MV-LK) model extends the previous model by allowing for heterogeneous genetic variances, providing environment-specific rates of shrinkage for a given management.
The number of parameters thus increases to $pq$ variances for all E$\times$M combinations and $up/2$ correlations between the different managements:
\begin{equation*}
	\boldsymbol{\kappa} = \left[\sigma_{1}^2, \sigma_{2}^2, \dots, \sigma_{pq}^2, \rho_{12}, \rho_{13}, \dots, \rho_{up}\right] \text{.}
\end{equation*}
The partial derivatives with respect to the variances are given by
\begin{equation}\label{eq:d_v_mvlk}
	\dfrac{\partial\Big[\s\s^\top\circ\left(\C_M\otimes\K_E\right)\Big]}{\partial \sigma_{l}^2} = \dfrac{\partial \s\s^\top}{\partial \sigma_{l}^2}\circ\left(\C_M\otimes\K_E\right) = \dfrac{\s\s^\top \circ \left(\mathbf{A}_{l} + \mathbf{A}_{l}^\top\right)}{\sigma_{l}^2} \circ \left(\C_M \otimes \K_E\right) \text{,}
\end{equation}
where $\mathbf{A}_{l} \in \mathbb{R}^{pq\times pq}$ now contains $0.5$ in the row corresponding to E$\times$M combination $l$.
The derivatives with respect to the correlations can be obtained by
\begin{equation}\label{eq:d_r_mvlk}
	\dfrac{\partial\Big[\mathbf{ss}^\top\circ\left(\C_M\otimes\K_E\right)\Big]}{\partial \rho_{ii'}} = \mathbf{ss}^\top\circ\left(\dfrac{\partial \C_M}{\partial \rho_{ii'}}\otimes\K_E\right) = \mathbf{ss}^\top \circ \left(\mathbf{B}_{ii'} \otimes \K_E\right) \text{.}
\end{equation}

\subsubsection*{Single variance Gaussian kernel}
The parameters and derivatives for the SV-GK model are identical to those given in (\ref{eq:d_v_svlk}) and (\ref{eq:d_r_svlk}), except for the fact that $\K_E$ is replaced by $e^{-h\D_E}$ and there is the additional parameter $h$ for which the derivative is
\begin{equation*}
	\dfrac{\partial\left(\Covmat_M\otimes e^{-h\D_E}\right)}{\partial h} = \Covmat_M\otimes \dfrac{\partial e^{-h\D_E}}{\partial h} = \Covmat_M \otimes \left(-\D_E \circ e^{-h\D_E}\right) \text{.}
\end{equation*}

\subsubsection*{Multiple variance Gaussian kernel}
Similarly, the derivatives with respect to the variances and correlations for the MV-GK model can be obtained by replacing $\K_E$ by $e^{-h\D_E}$ in (\ref{eq:d_v_mvlk}) and (\ref{eq:d_r_mvlk}).
The derivative with respect to the bandwidth is then
\begin{equation*}
	\dfrac{\partial\Big[\mathbf{ss}^\top\circ\left(\C_M\otimes e^{-h\D_E}\right)\Big]}{\partial h} = \mathbf{ss}^\top\circ\left(\C_M\otimes \dfrac{\partial e^{-h\D_E}}{\partial h}\right) = \mathbf{ss}^\top \circ \Big[\C_M \otimes \left(-\D_E \circ e^{-h\D_E}\right)\Big] \text{.}
\end{equation*}

\subsubsection*{Multiple bandwidth single variance Gaussian kernel}
The MB-SV-GK model has multiple bandwidths corresponding to each level of management as well as the correlations between managements.
Derivatives with respect to variances can be obtained as
\begin{equation*}
	\dfrac{\partial \Big[\left(\Covmat_M \otimes \mathbf{J}_q\right) \circ e^{-\mathbf{H}\otimes\D_E}\Big]}{\partial \sigma^2_i} = \left(\dfrac{\partial \Covmat_M}{\partial \sigma^2_i} \otimes \mathbf{J}_q\right) \circ e^{-\mathbf{H}\otimes\D_E} = \left[\dfrac{\Covmat_M \circ \left(\mathbf{A}_i + \mathbf{A}_i^\top\right)}{\sigma^2_i} \otimes \mathbf{J}_q\right] \circ e^{-\mathbf{H} \otimes \D_E} \text{,}
\end{equation*}
derivatives with respect to correlations as
\begin{equation*}
	\dfrac{\partial \Big[\left(\Covmat_M \otimes \mathbf{J}_q\right) \circ e^{-\mathbf{H}\otimes\D_E}\Big]}{\partial \rho_{ii'}} = \left(\dfrac{\partial \Covmat_M}{\partial \rho_{ii'}} \otimes \mathbf{J}_q\right) \circ e^{-\mathbf{H}\otimes\D_E} = \Big[\left(\mathbf{s}_M\mathbf{s}_M^\top \circ \mathbf{B}_{ii'}\right) \otimes \mathbf{J}_q\Big] \circ e^{-\mathbf{H}\otimes\D_E} \text{,}
\end{equation*}
and derivatives with respect to one of the bandwidths as
\begin{equation*}
	\dfrac{\partial \Big[\left(\Covmat_M \otimes \mathbf{J}_q\right) \circ e^{-\mathbf{H}\otimes\D_E}\Big]}{\partial h_{ii'}} = \left(\Covmat_M \otimes \mathbf{J}_q\right) \circ \dfrac{\partial e^{-\mathbf{H}\otimes\D_E}}{\partial h_{ii'}} = \left(\Covmat_M \otimes \mathbf{J}_q\right) \circ \left(\mathbf{B}_{ii'} \otimes -\D_E\right) \circ e^{-\mathbf{H}\otimes\D_E} \text{,}
\end{equation*}
where $\mathbf{B}_{ii'}$ is as described before, except that for bandwidths $i$ can be the same as $i'$.

\subsubsection*{Multiple bandwidth multiple variance Gaussian kernel}
Derivatives for the MB-MV-GK model are similar.
For variances:
\begin{equation*}
	\dfrac{\partial \Big[\s\s^{\top} \circ \left(\C_M \otimes \mathbf{J}_q\right) \circ e^{-\mathbf{H}\otimes\D_E}\Big]}{\partial \sigma^2_l} = \dfrac{\partial \s\s^{\top}}{\partial \sigma^2_l} \circ \left(\C_M \otimes \mathbf{J}_q\right) \circ e^{-\mathbf{H}\otimes\D_E} = \dfrac{\s\s^{\top} \circ \left(\mathbf{A}_{l} + \mathbf{A}_{l}^\top\right)}{\sigma^2_l} \circ \left(\C_M \otimes \mathbf{J}_q\right) \circ e^{-\mathbf{H}\otimes\D_E} \text{,}
\end{equation*}
for correlations:
\begin{equation*}
	\dfrac{\partial \Big[\s\s^{\top} \circ \left(\C_M \otimes \mathbf{J}_q\right) \circ e^{-\mathbf{H}\otimes\D_E}\Big]}{\partial \rho_{ii'}} = \s\s^{\top} \circ \left(\dfrac{\partial \C_M}{\partial \rho_{ii'}} \otimes \mathbf{J}_q\right) \circ e^{-\mathbf{H}\otimes\D_E} = \s\s^\top \circ \left(\mathbf{B}_{ii'} \otimes \mathbf{J}_q\right) \circ e^{-\mathbf{H}\otimes\D_E} \text{,}
\end{equation*}
and for bandwidths:
\begin{equation*}
	\dfrac{\partial \Big[\s\s^{\top} \circ \left(\C_M \otimes \mathbf{J}_q\right) \circ e^{-\mathbf{H}\otimes\D_E}\Big]}{\partial h_{ii'}} = \s\s^{\top} \circ \left(\C_M \otimes \mathbf{J}_q\right) \circ \dfrac{\partial e^{-\mathbf{H}\otimes\D_E}}{\partial h_{ii'}} = \s\s^{\top} \circ \left(\C_M \otimes \mathbf{J}_q\right) \circ \left(\mathbf{B}_{ii'} \otimes -\D_E\right) \circ e^{-\mathbf{H}\otimes\D_E} \text{.}
\end{equation*}


\newpage
\FloatBarrier
\begingroup
\setlength{\tabcolsep}{10pt} 
\renewcommand{\arraystretch}{2} 
\begin{table}[H]
	\caption{Overview of the models described in Section \ref{sec:models}. ADD = additive main effects, FA-$m$ = order $m$ factor analytic, SV-LK = single variance linear kernel, MV-LK = multiple variance linear kernel, SV-GK = single variance Gaussian kernel, MV-GK = multiple variance Gaussian kernel, MB-SV-GK = multiple bandwidth single variance Gaussian kernel, MB-MV-GK = multiple bandwidth multiple variance Gaussian kernel. Note that the $(m^2-m)/2$ constrained loadings have been subtracted for the FA models.}
	\label{model_overview}
	\begin{tabular}{ cccc }
		Model & $\mathbb{V}\text{ar}\left(\mathbf{u}\right)$ & Number of parameters for $\mathbb{V}\text{ar}\left(\mathbf{u}\right)$ \\
		\hline
		ADD & $\sigma^2 \K_G$ & $1$ \\
		FA-$m$ & $\left(\LA\LA^\top+\PS\right) \otimes \K_G$ & $pqm+pq-(m^2-m)/2$ \\
		SV-LK & $\Covmat_M \otimes \K_E \otimes \K_G$ & $\left(p^2+p\right)/2$ &  \\
		MV-LK & $\Big[\mathbf{s}\mathbf{s}^\top \circ \left(\C_M \otimes \K_E\right)\Big] \otimes \K_G$ & $pq + \left(p^2-p\right)/2$ \\
		SV-GK & $\Covmat_M \otimes e^{-h\D_E} \otimes \K_G$ & $\left(p^2+p\right)/2 + 1$ &  \\
		MV-GK & $\Big[\mathbf{s}\mathbf{s}^\top \circ \left(\C_M \otimes e^{-h\D_E}\right)\Big] \otimes \K_G$ & $pq + \left(p^2-p\right)/2 + 1$ \\
		MB-SV-GK & $\Big[\left(\Covmat_M \otimes \mathbf{J}_q\right) \circ e^{-\mathbf{H}\otimes\D_E}\Big] \otimes \K_G$ & $p + p^2$ \\
		MB-MV-GK & $\s\s^{\top} \circ \Big[\left(\C_M \otimes \mathbf{J}_q\right) \circ e^{-\mathbf{H}\otimes\D_E}\Big] \otimes \K_G$ & $pq + p^2$
	\end{tabular}
\end{table}
\endgroup
\newpage
\begin{center}
	\begin{minipage}{\linewidth}
		\centering
		\textbf{\large Supplementary Material: REML implementations of kernel-based genomic prediction models for genotype $\times$ environment $\times$ management interactions}
		\vskip0.5cm
		Killian A.C.\ Melsen$^{1*}$, Salvador A. Gezan$^2$, Daniel J.\ Tolhurst$^3$, 
		Fred A.\ van Eeuwijk$^1$, Carel F.W.\ Peeters$^1$
	\end{minipage}
	\vskip0.5cm
	\begin{minipage}{\linewidth}
		\raggedright
		$^1$Mathematical \& Statistical Methods group - Biometris, Wageningen University \& Research, PO Box 16, 6700 AA, Wageningen, The Netherlands\\
		$^2$VSN International, 2 Amberside House, Wood Lane, HP2 4TP, Hemel Hempstead, United Kingdom\\
		$^3$The Roslin Institute and Royal (Dick) School of Veterinary Science, University of Edinburgh, Easter Bush, Midlothian, EH25 9RG, United Kingdom\\
		$^*$ \url{killian.melsen@wur.nl}
	\end{minipage}
	\vskip2cm
\end{center}
\setcounter{equation}{0}
\setcounter{figure}{0}
\setcounter{table}{0}
\setcounter{page}{1}
\setcounter{section}{0}
\makeatletter
\renewcommand{\theequation}{S\arabic{equation}}
\renewcommand{\thefigure}{S\arabic{figure}}
\renewcommand{\bibnumfmt}[1]{[S#1]}
\renewcommand{\citenumfont}[1]{S#1}
\renewcommand{\thetable}{S\arabic{table}}
\renewcommand{\thesection}{S.\arabic{section}}
\begin{table}[h]
	\caption{Overview of the geographical coordinates, sowing, and harvest dates (yyyy-mm-dd) to obtain environmental covariables for the $14$ environments in the BRIWECS dataset.
	The environmental covariables used for the kernels included RH2M, WS2M, PRECTOT, EVPTRNS, P\_ETP, VPD, ALLSKY\_SFC\_LW\_DWN, ALLSKY\_SFC\_SW\_DWN, ALLSKY\_SFC\_SW\_DNI, ALLSKY\_SFC\_PAR\_TOT, ALLSKY\_SFC\_UVA, ALLSKY\_SFC\_UVB, RTA, n, N, T2M, T2M\_MAX, and T2M\_MIN.
	See the documentation of the {\ttfamily EnvRtype::get\_weather()} function for details on these covariables.}
	\label{SM:BRIWECS_table}
	\begin{tabular}{ ccccc }
		Environment & Latitude & Longitude & Sowing date & Harvest date \\
		\hline
		HAN\_2015 & 52.24382 & 9.818108 & 2014-10-28 & 2015-07-22 \\
		HAN\_2016 & 52.24382 & 9.818108 & 2015-11-04 & 2016-07-10 \\
		HAN\_2017 & 52.24382 & 9.818108 & 2016-11-01 & 2017-07-09 \\
		
		KAL\_2015 & 50.61310 & 6.994236 & 2014-10-28 & 2015-07-13 \\
		KAL\_2016 & 50.61310 & 6.994236 & 2015-10-28 & 2016-07-21 \\
		KAL\_2017 & 50.61310 & 6.994236 & 2016-10-26 & 2017-07-08 \\
		
		KIE\_2015 & 54.31568 & 9.980414 & 2014-10-23 & 2015-07-13 \\
		KIE\_2016 & 54.31568 & 9.980414 & 2015-10-02 & 2016-06-21 \\
		KIE\_2017 & 54.31568 & 9.980414 & 2016-09-21 & 2017-07-22 \\
		
		QLB\_2015 & 51.76921 & 11.145669 & 2014-10-18 & 2015-07-07 \\
		QLB\_2016 & 51.76921 & 11.145669 & 2015-10-10 & 2016-07-07 \\
		QLB\_2017 & 51.76921 & 11.145669 & 2016-11-02 & 2017-07-10 \\
		
		RHH\_2015 & 50.76050 & 8.875599 & 2014-10-28 & 2015-07-18 \\
		RHH\_2016 & 50.76050 & 8.875599 & 2015-10-20 & 2016-07-09 \\
		
	\end{tabular}
\end{table}

\begin{table}[h]
	\caption{BRIWECS and DROPS narrow-sense heritabilities for all combinations of managements and models using (latent) environmental covariables. Heritabilities were computed using variance components averaged over environments.}
	\label{SM:h2_table}
	\begin{tabular}{ cccc }
		Dataset & Management & Model & $h^2$ \\
		\hline
		BRIWECS & HN & FVP   	& 0.88 \\
		 		& 	 & ADD   	& 0.88 \\
		 		& 	 & FA-1   	& 0.90 \\
			    &    & FA-2   	& 0.90 \\
			    &    & FA-3   	& 0.90 \\
			    &    & SV-LK  	& 0.90 \\
			    &    & SV-GK  	& 0.91 \\
			    &    & MV-LK  	& 0.89 \\
			    &    & MV-GK    & 0.90 \\
			    &    & MB-SV-LK & 0.90 \\
			    &    & MB-MV-GK & 0.90 \\
		BRIWECS & LN & FVP   	& 0.88 \\
		 		&  	 & ADD   	& 0.84 \\
	     		&  	 & FA-1   	& 0.84 \\
	 		    &    & FA-2   	& 0.84 \\
	 		    &    & FA-3   	& 0.84 \\
	 		    &    & SV-LK  	& 0.83 \\
	 		    &    & SV-GK  	& 0.84 \\
	 		    &    & MV-LK  	& 0.82 \\
	 		    &    & MV-GK    & 0.84 \\
	 		    &    & MB-SV-LK & 0.83 \\
	 		    &    & MB-MV-GK & 0.84 \\
	 	DROPS   & W  & FVP   	& 0.64 \\
	 			&    & ADD   	& 0.64 \\
	 			&    & FA-1   	& 0.66 \\
	 			&    & FA-2   	& 0.66 \\
	 			&    & FA-3   	& 0.66 \\
	 			&    & SV-LK  	& 0.65 \\
	 			&    & SV-GK  	& 0.68 \\
	 			&    & MV-LK  	& 0.64 \\
	 			&    & MV-GK    & 0.67 \\
	 			&    & MB-SV-LK & 0.67 \\
	 			&    & MB-MV-GK & 0.67 \\
	 	DROPS   & R  & FVP   	& 0.78 \\
	 	   		&    & ADD   	& 0.76 \\
	       		&    & FA-1   	& 0.76 \\
	  			&    & FA-2   	& 0.76 \\
	  			&    & FA-3   	& 0.76 \\
	  			&    & SV-LK  	& 0.75 \\
	  			&    & SV-GK  	& 0.77 \\
	  			&    & MV-LK  	& 0.74 \\
	  			&    & MV-GK    & 0.76 \\
	  			&    & MB-SV-LK & 0.77 \\
	  			&    & MB-MV-GK & 0.76 \\
	\end{tabular}
\end{table}

\begin{figure}[h]
	\includegraphics[width=0.85\textwidth]{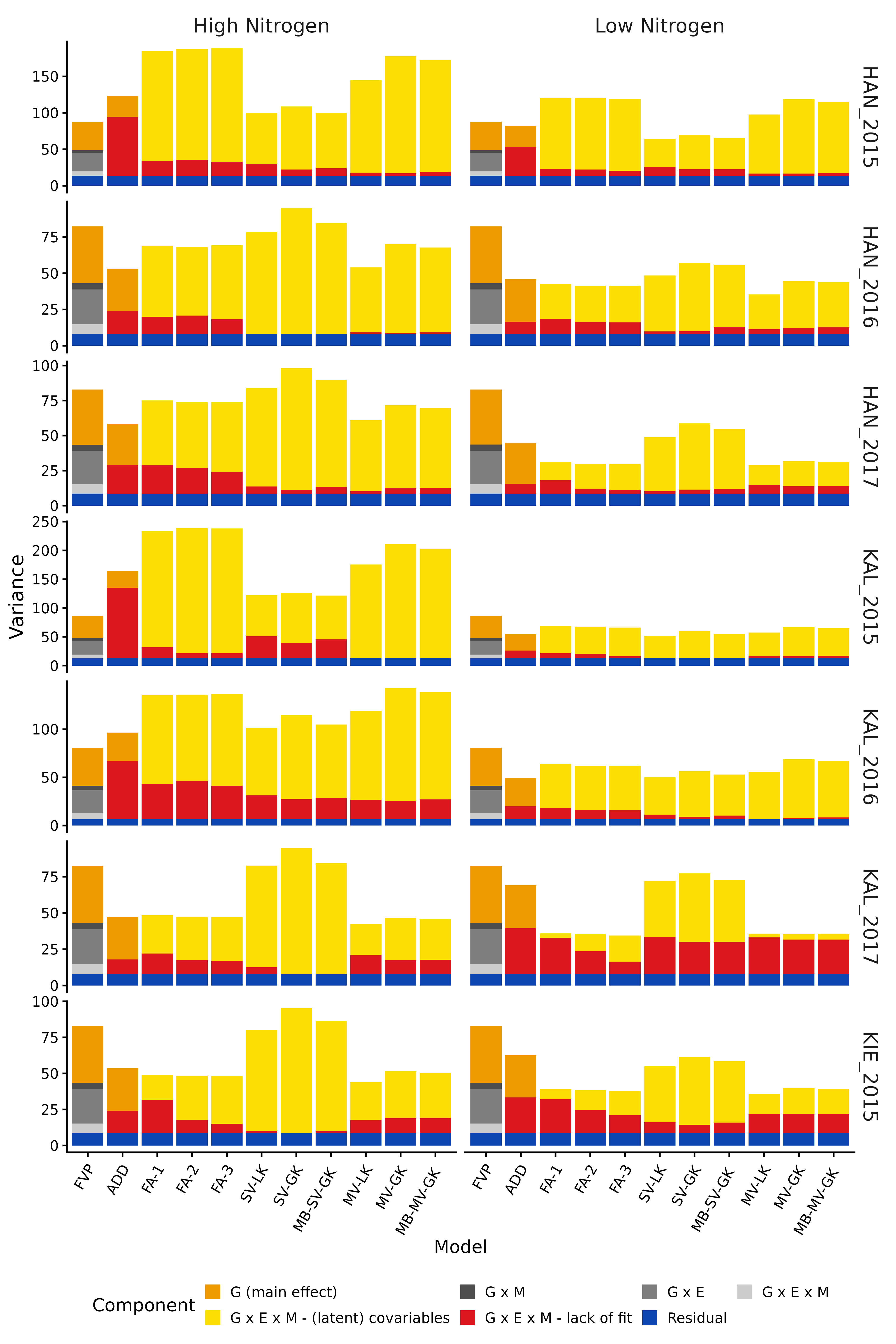}
	\caption{Partitioning of G$\times$E$\times$M variance into structured, lack of fit, and residual effects for each environment in the BRIWECS dataset.
	\\[5pt]
	\textbf{Alt text:} Bar charts showing the amounts of variance explained by the different effects in the variance partitioning models for the BRIWECS dataset for the first seven environments.}
	\label{fig:BRIWECS_LOF_EXTRA_A}
\end{figure}
\begin{figure}[h]
	\includegraphics[width=0.85\textwidth]{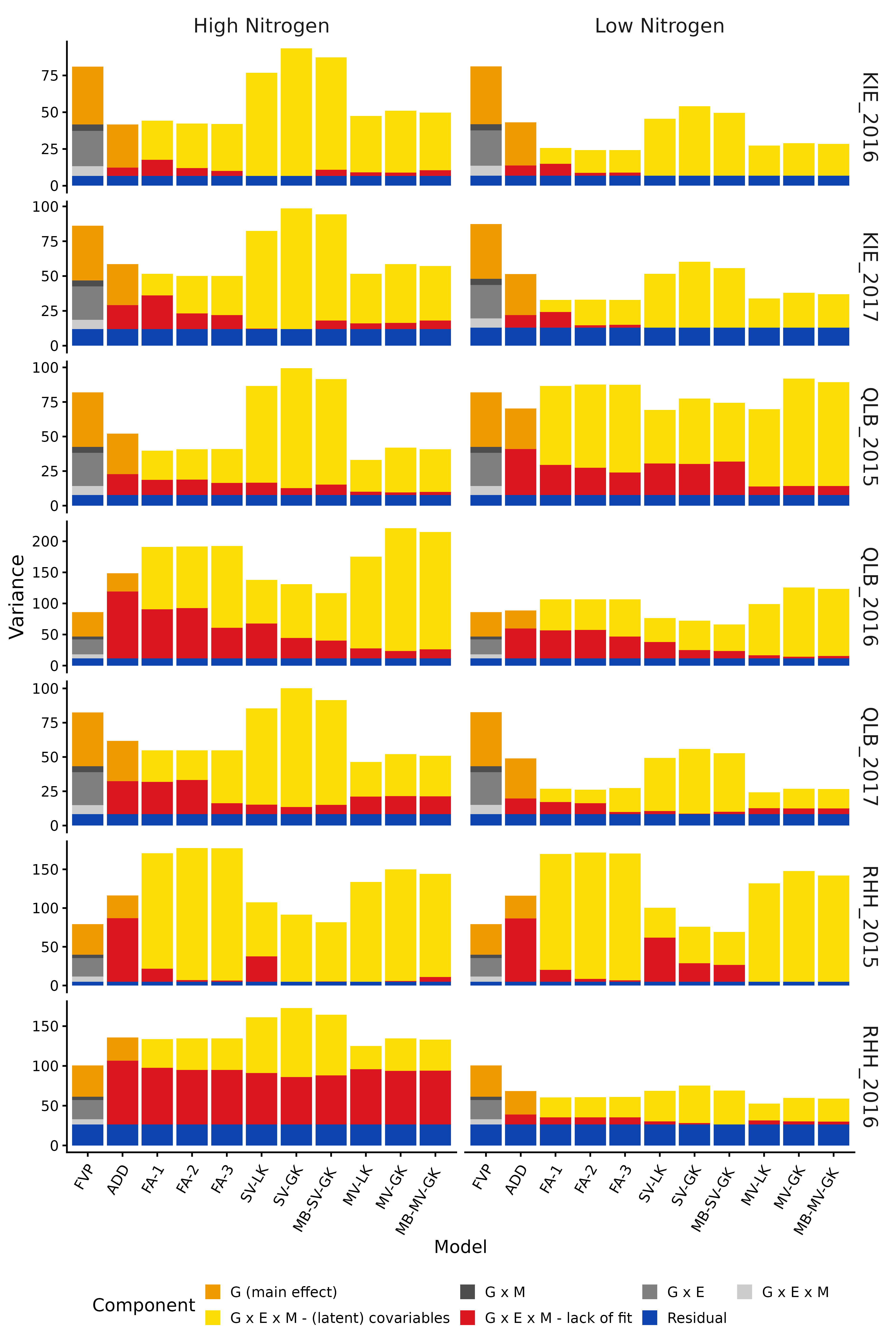}
	\caption{Continued. Partitioning of G$\times$E$\times$M variance into structured, lack of fit, and residual effects for each environment in the BRIWECS dataset.
	\\[5pt]
	\textbf{Alt text:} Bar charts showing the amounts of variance explained by the different effects in the variance partitioning models for the BRIWECS dataset for the remaining environments.}
	\label{fig:BRIWECS_LOF_EXTRA_B}
\end{figure}

\begin{figure}[h]
	\includegraphics[width=0.85\textwidth]{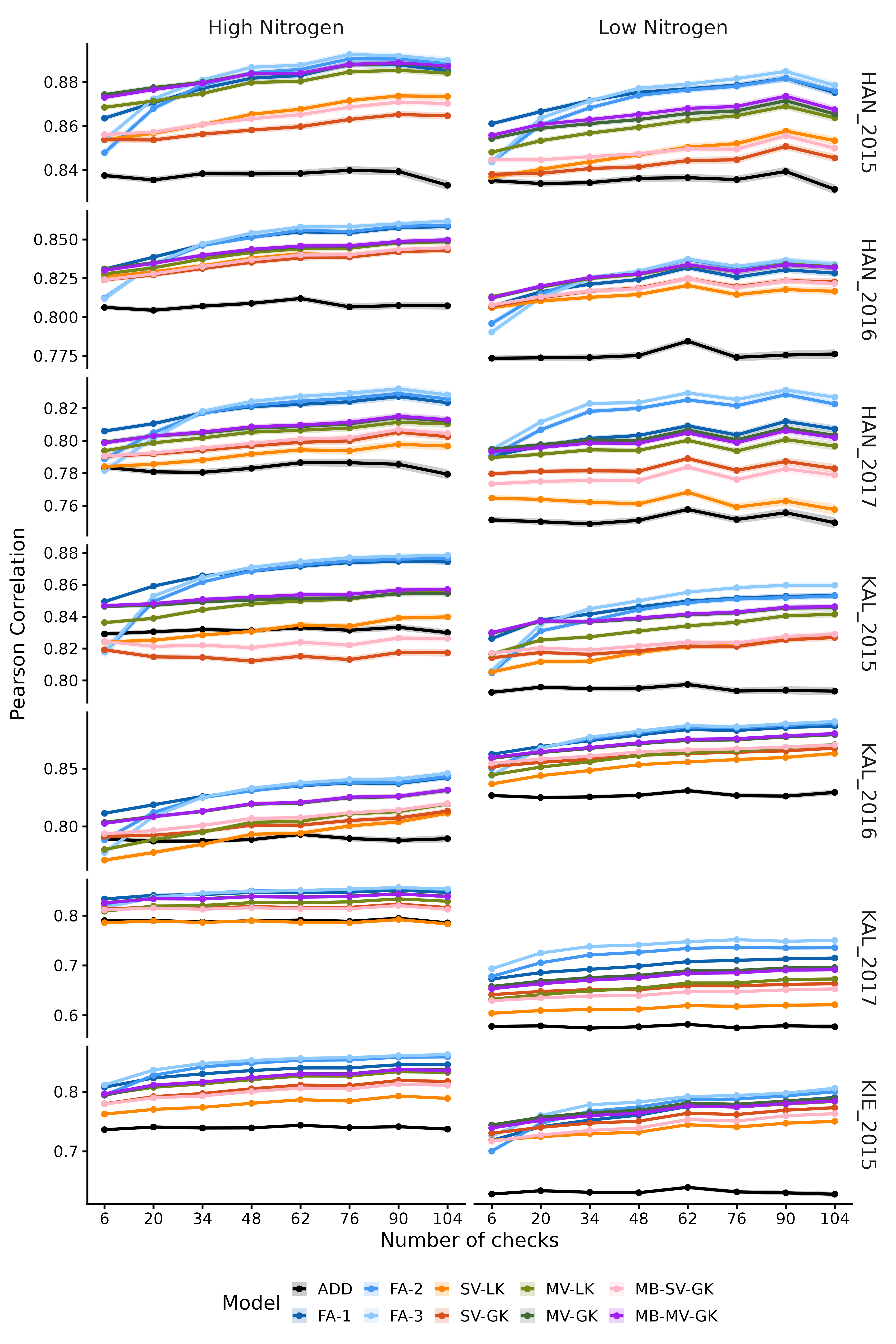}
	\caption{Pearson correlation between BLUPs from the LMMs and centered test-set phenotypes for the BRIWECS dataset for each environment. Shaded areas represent standard errors.
	\\[5pt]
	\textbf{Alt text:} Line graphs showing the genomic prediction accuracies (Pearson correlation) obtained using the different models and levels of sparsity for the first seven environments of the BRIWECS dataset.}
	\label{fig:BRIWECS_CORR_EXTRA_A}
\end{figure}
\begin{figure}[h]
	\includegraphics[width=0.85\textwidth]{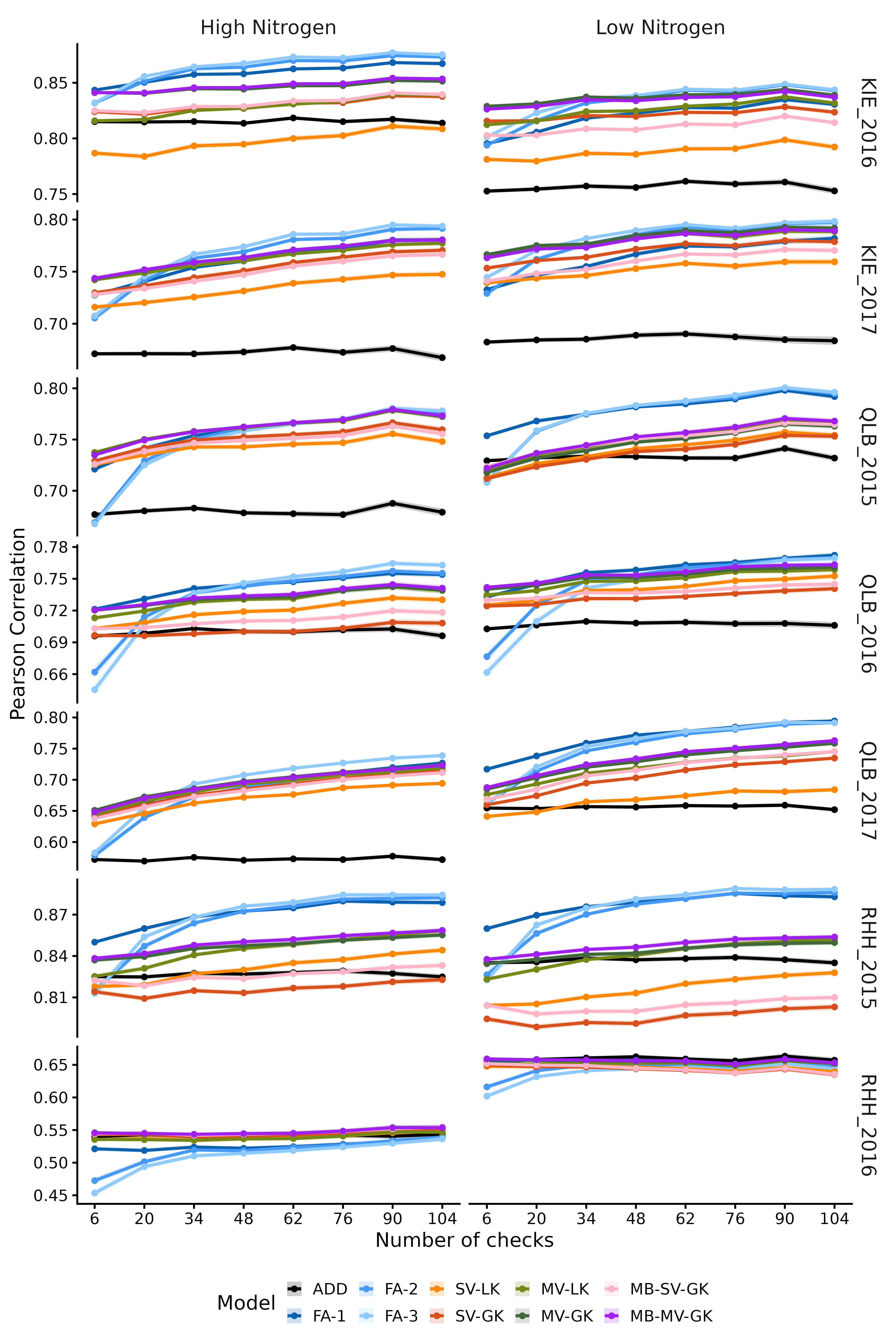}
	\caption{Continued. Pearson correlation between BLUPs from the LMMs and centered test-set phenotypes for the BRIWECS dataset for each environment. Shaded areas represent standard errors.
	\\[5pt]
	\textbf{Alt text:} Line graphs showing the genomic prediction accuracies (Pearson correlation) obtained using the different models and levels of sparsity for the remaining environments of the BRIWECS dataset.}
	\label{fig:BRIWECS_CORR_EXTRA_B}
\end{figure}
\begin{figure}[h]
	\includegraphics[width=0.85\textwidth]{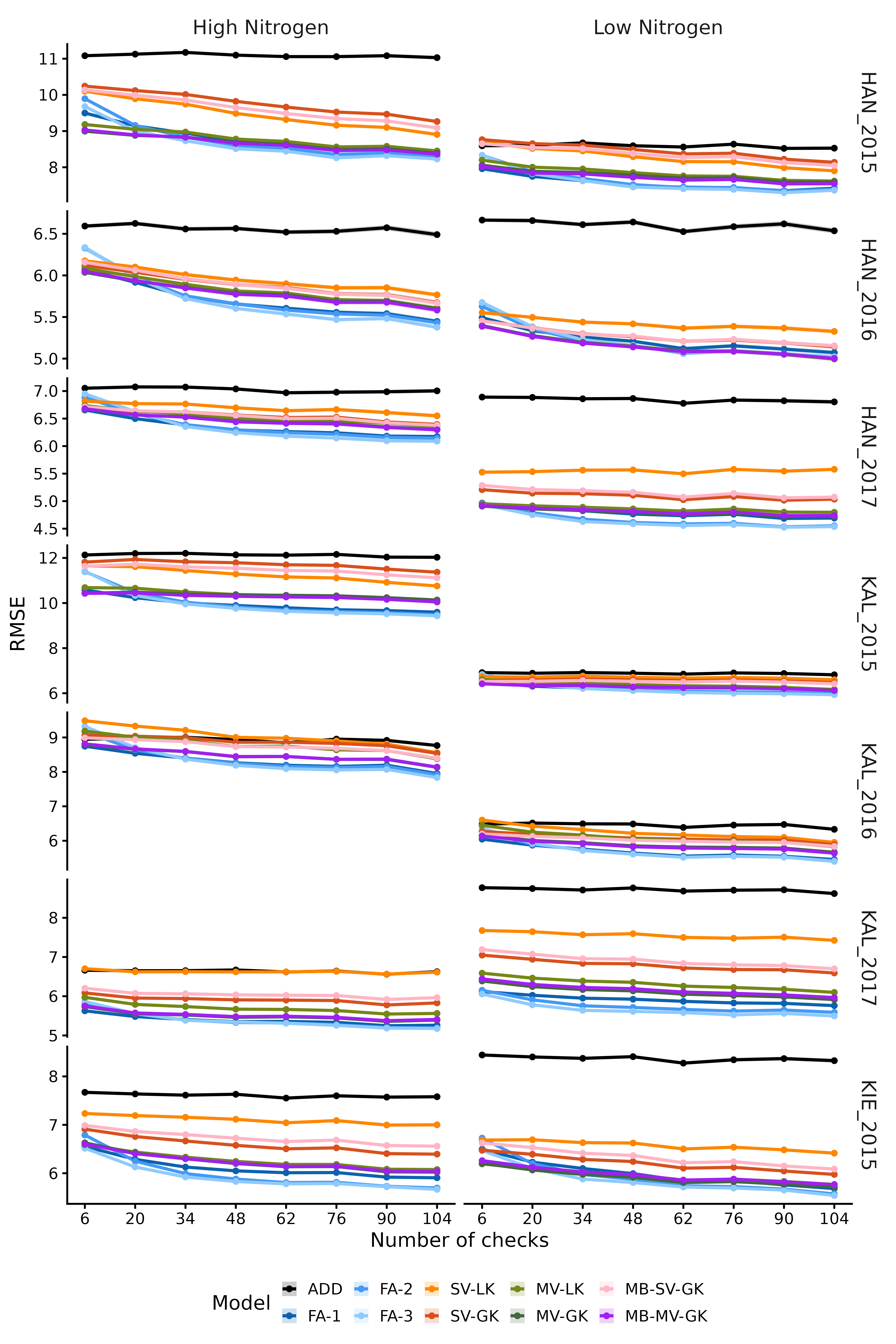}
	\caption{RMSE between BLUPs from the LMMs and centered test-set phenotypes for the BRIWECS dataset for each environment. Shaded areas represent standard errors.
	\\[5pt]
	\textbf{Alt text:} Line graphs showing the genomic prediction accuracies (RMSE) obtained using the different models and levels of sparsity for the first seven environments of the BRIWECS dataset.}
	\label{fig:BRIWECS_RMSE_EXTRA_A}
\end{figure}
\begin{figure}[h]
	\includegraphics[width=0.85\textwidth]{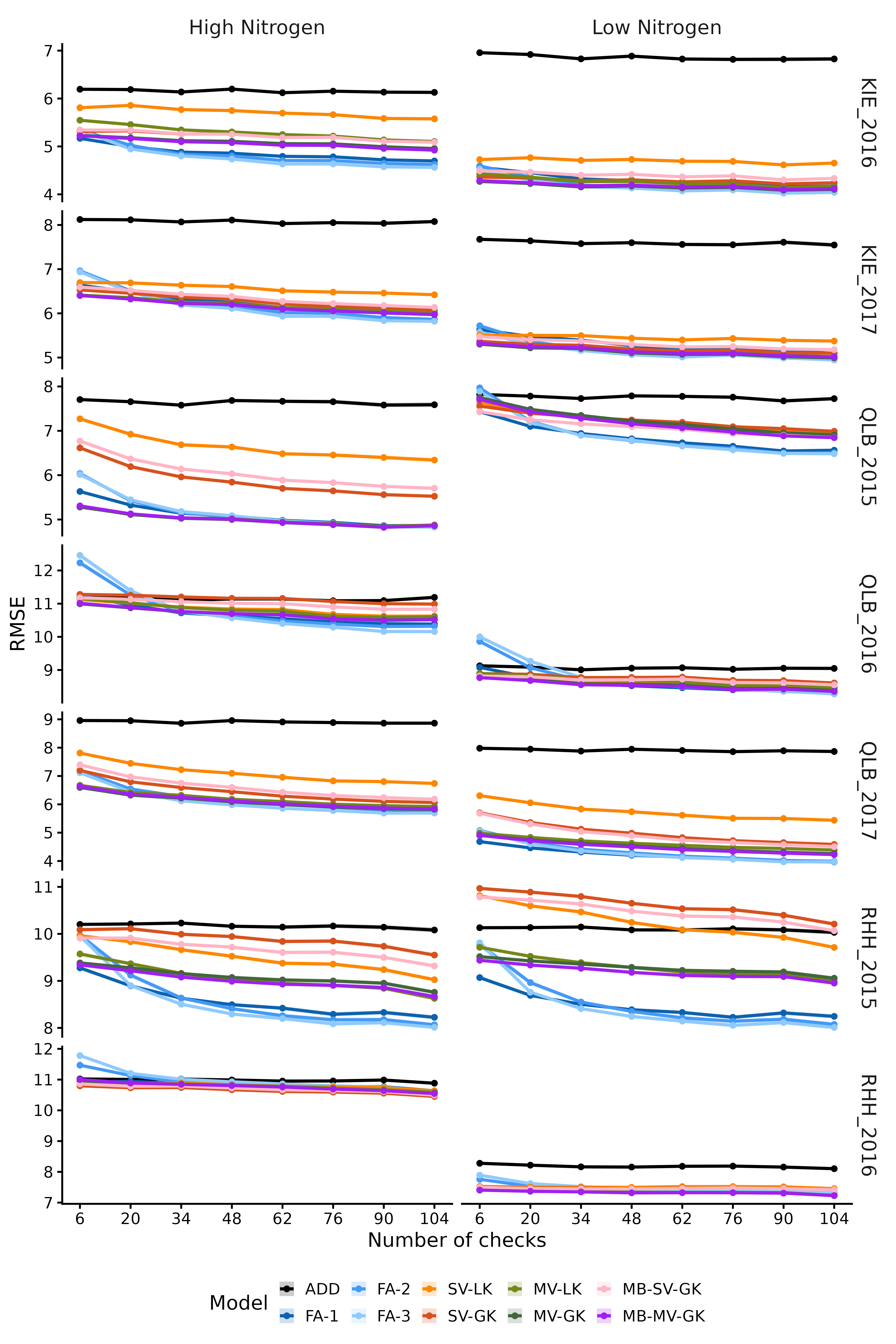}
	\caption{Continued. RMSE between BLUPs from the LMMs and centered test-set phenotypes for the BRIWECS dataset for each environment. Shaded areas represent standard errors.
	\\[5pt]
	\textbf{Alt text:} Line graphs showing the genomic prediction accuracies (RMSE) obtained using the different models and levels of sparsity for the remaining environments of the BRIWECS dataset.}
	\label{fig:BRIWECS_RMSE_EXTRA_B}
\end{figure}

\begin{figure}[h]
	\includegraphics[width=0.85\textwidth]{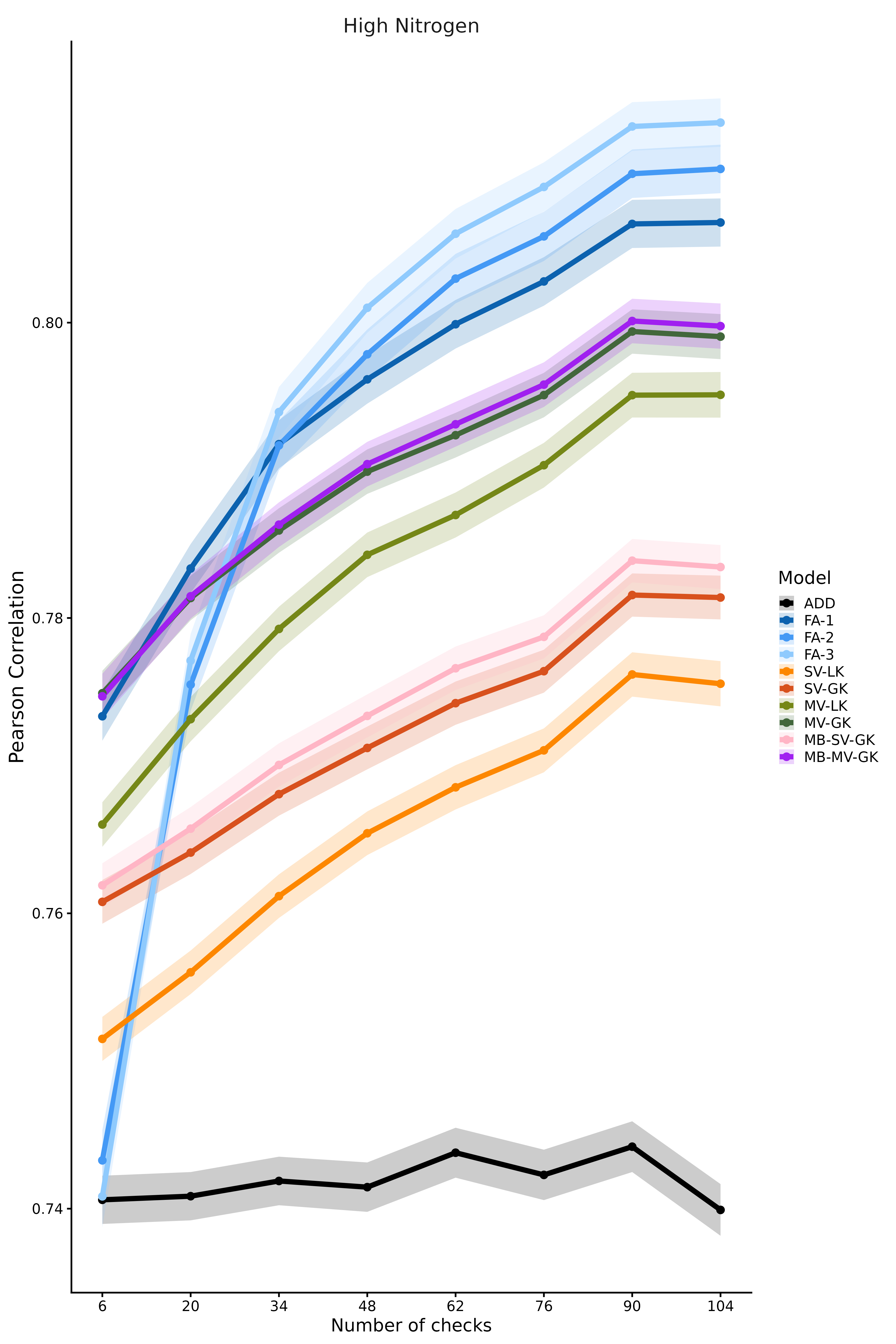}
	\caption{An enlarged version of the Pearson correlation, high nitrogen panel of Figure \ref{fig:BRIWECS_CV2}.
		\\[5pt]
		\textbf{Alt text:} An enlarged version of the Pearson correlation, high nitrogen panel of Figure \ref{fig:BRIWECS_CV2}.}
	\label{fig:BRIWECS_CV2_PCHN}
\end{figure}
\begin{figure}[h]
	\includegraphics[width=0.85\textwidth]{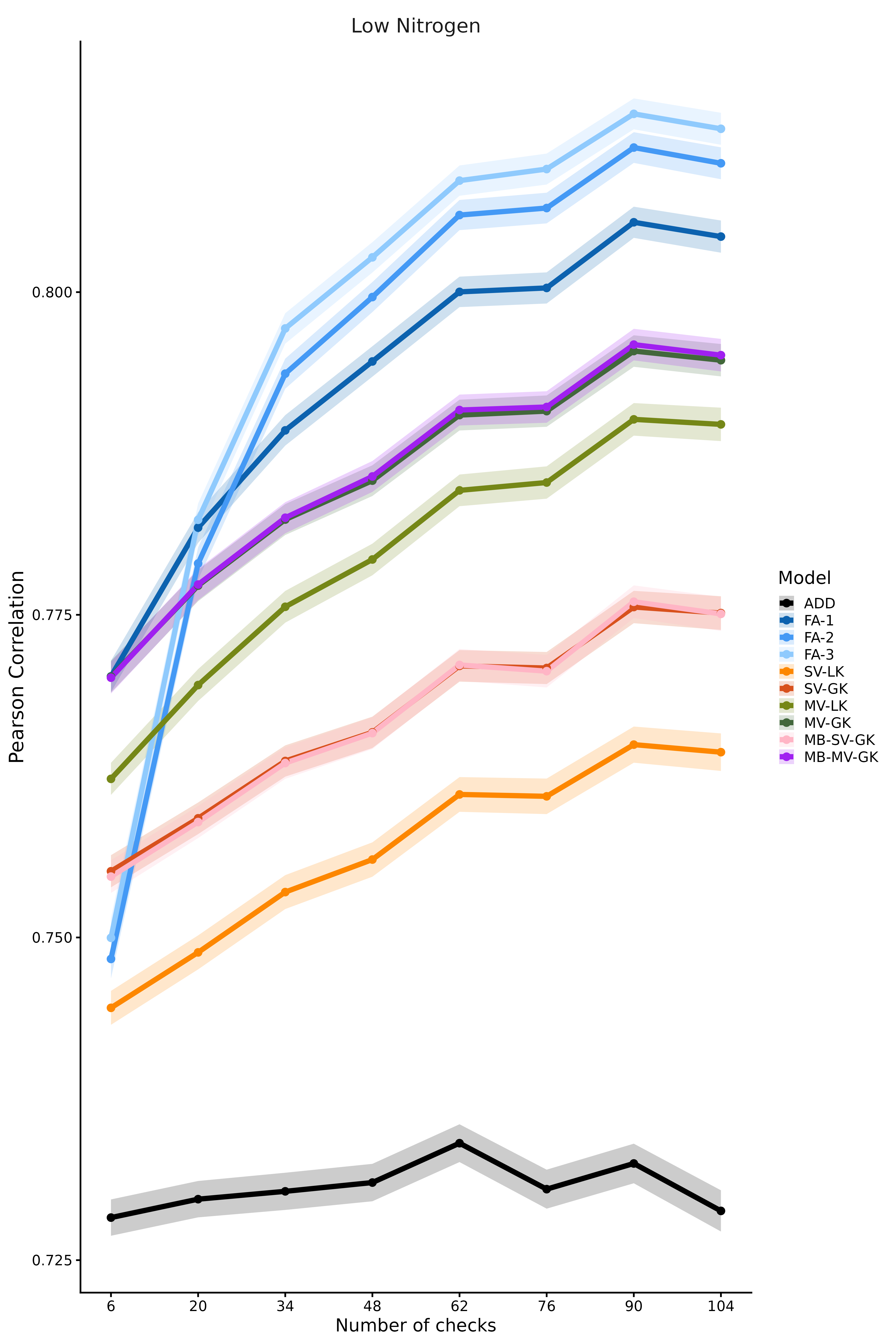}
	\caption{An enlarged version of the Pearson correlation, low nitrogen panel of Figure \ref{fig:BRIWECS_CV2}.
		\\[5pt]
		\textbf{Alt text:} An enlarged version of the Pearson correlation, low nitrogen panel of Figure \ref{fig:BRIWECS_CV2}.}
	\label{fig:BRIWECS_CV2_PCLN}
\end{figure}
\begin{figure}[h]
	\includegraphics[width=0.85\textwidth]{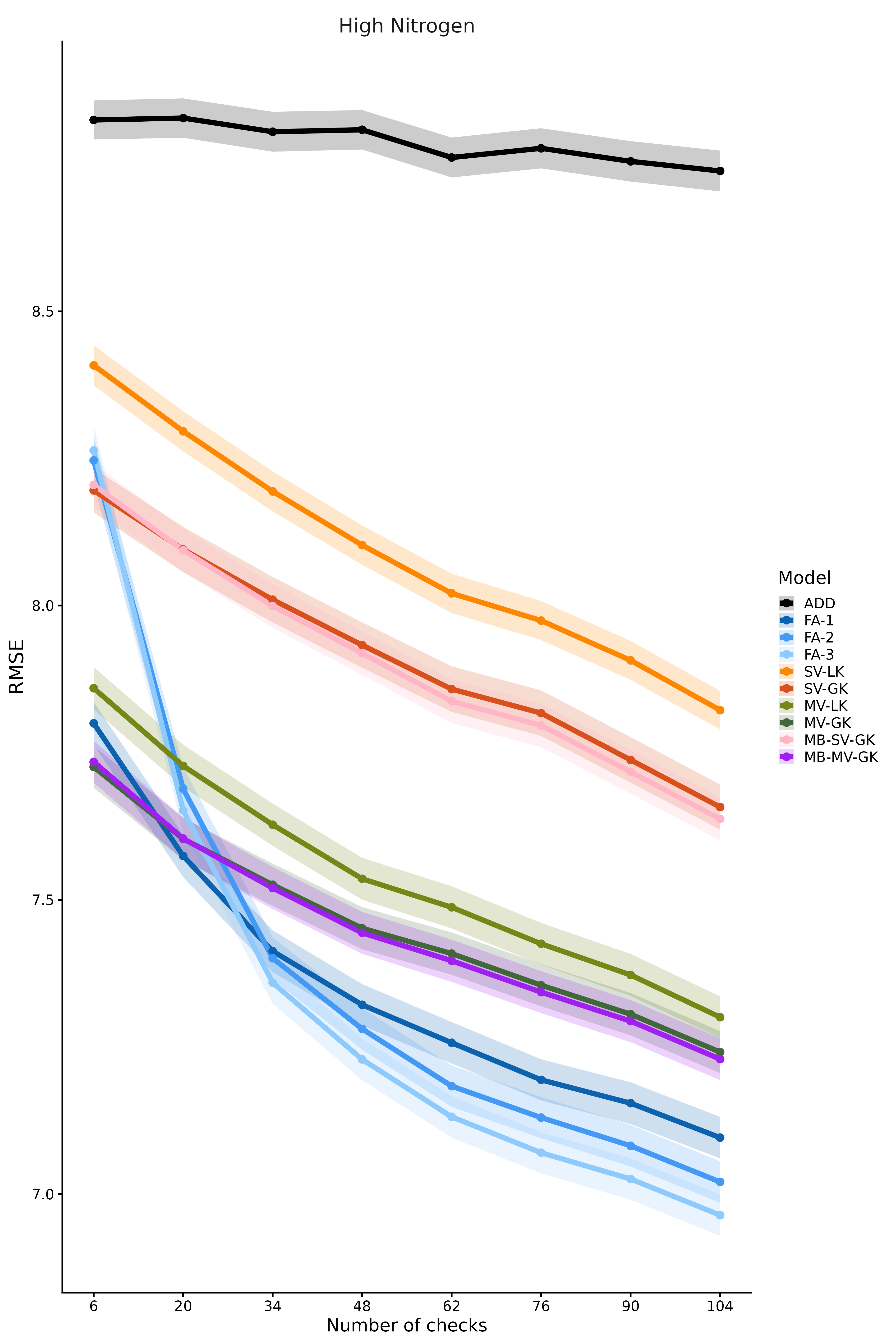}
	\caption{An enlarged version of the RMSE, high nitrogen panel of Figure \ref{fig:BRIWECS_CV2}.
		\\[5pt]
		\textbf{Alt text:} An enlarged version of the RMSE, high nitrogen panel of Figure \ref{fig:BRIWECS_CV2}.}
	\label{fig:BRIWECS_CV2_RMSEHN}
\end{figure}
\begin{figure}[h]
	\includegraphics[width=0.85\textwidth]{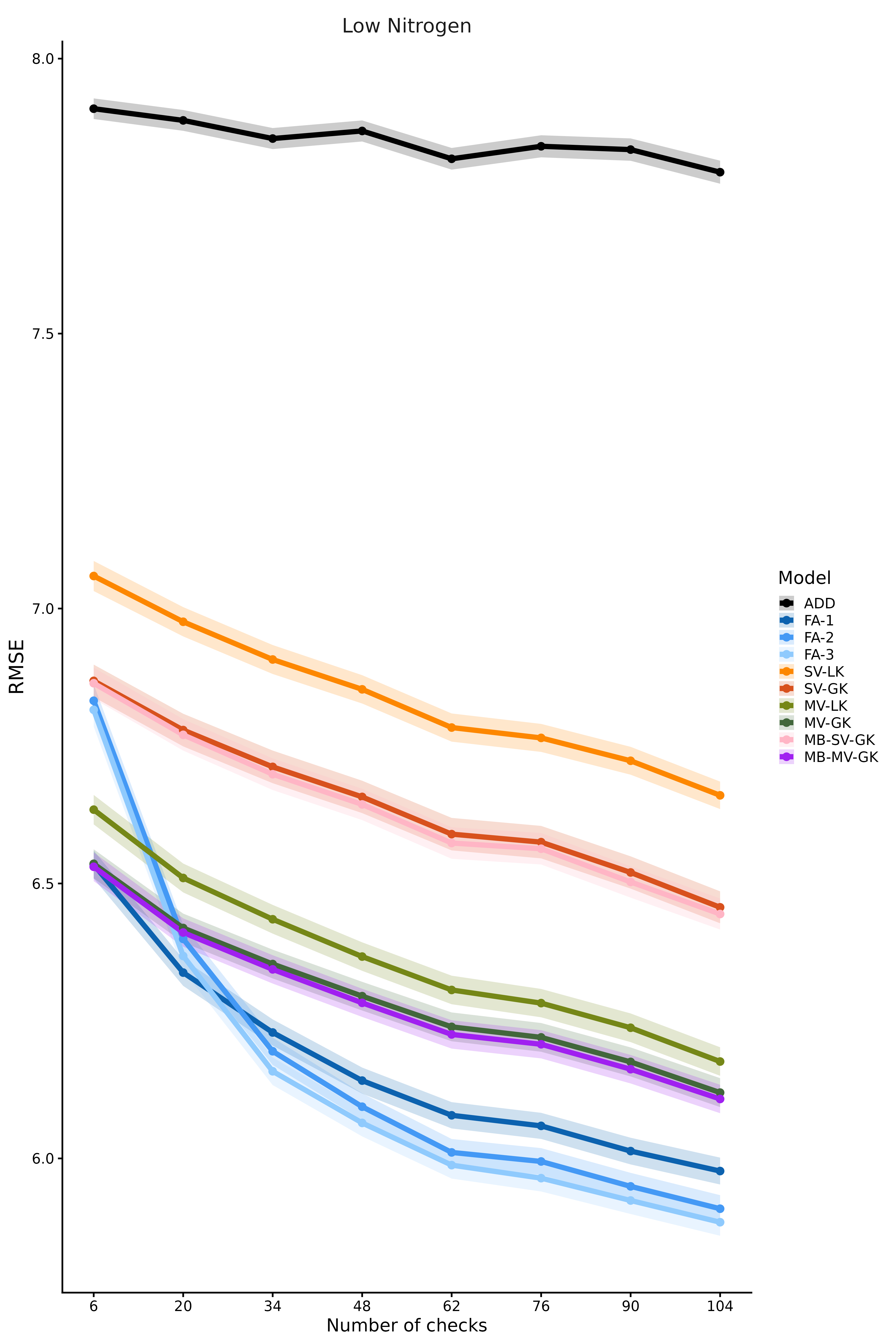}
	\caption{An enlarged version of the RMSE, low nitrogen panel of Figure \ref{fig:BRIWECS_CV2}.
		\\[5pt]
		\textbf{Alt text:} An enlarged version of the RMSE, low nitrogen panel of Figure \ref{fig:BRIWECS_CV2}.}
	\label{fig:BRIWECS_CV2_RMSELN}
\end{figure}

\begin{figure}[h]
	\includegraphics[width=0.85\textwidth]{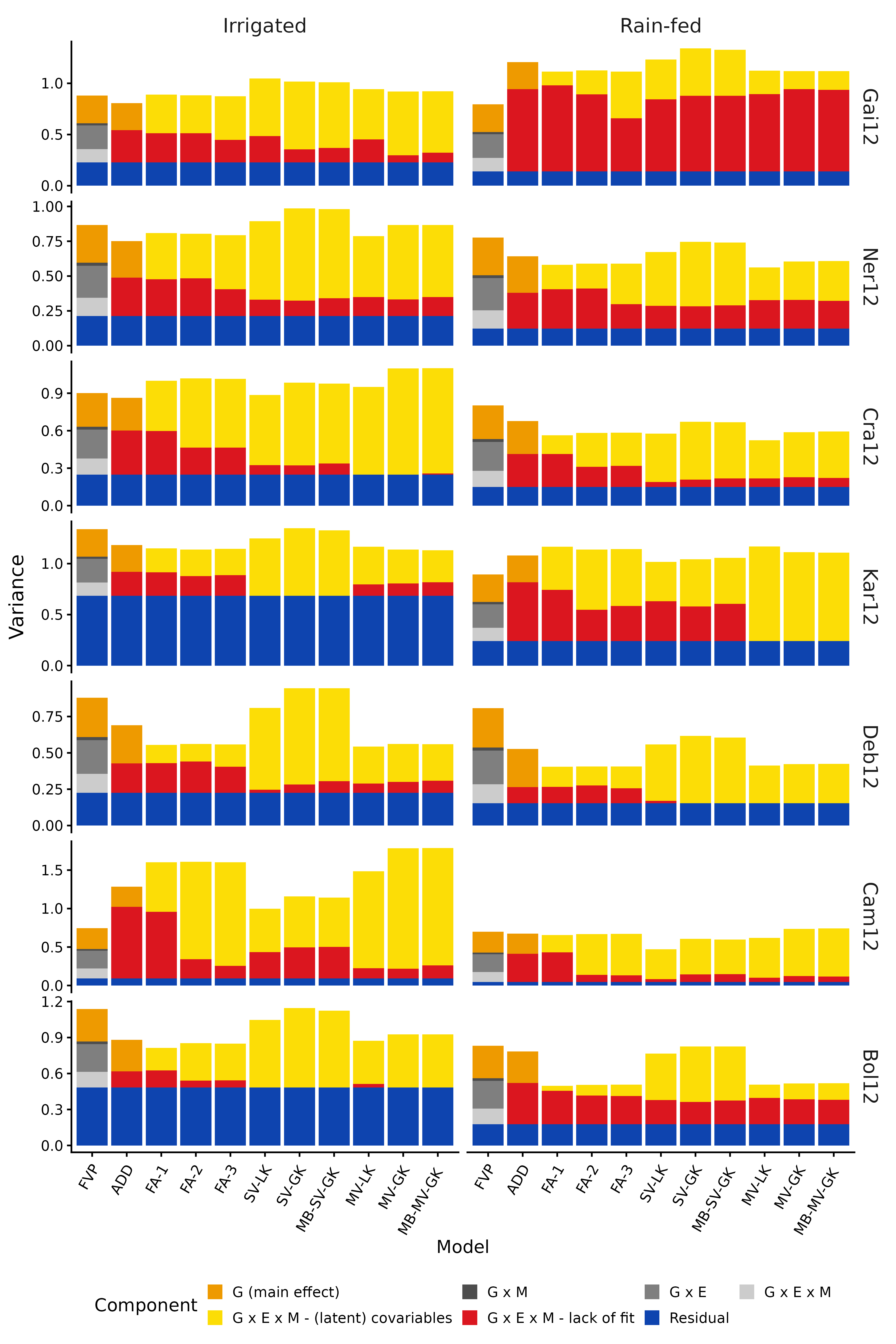}
	\caption{Partitioning of G$\times$E$\times$M variance into structured, lack of fit, and residual effects for each environment in the DROPS dataset.
	\\[5pt]
	\textbf{Alt text:} Bar charts showing the amounts of variance explained by the different effects in the variance partitioning models for the DROPS dataset for the first seven environments.}
	\label{fig:DROPS_LOF_EXTRA_A}
\end{figure}
\begin{figure}[h]
	\includegraphics[width=0.85\textwidth]{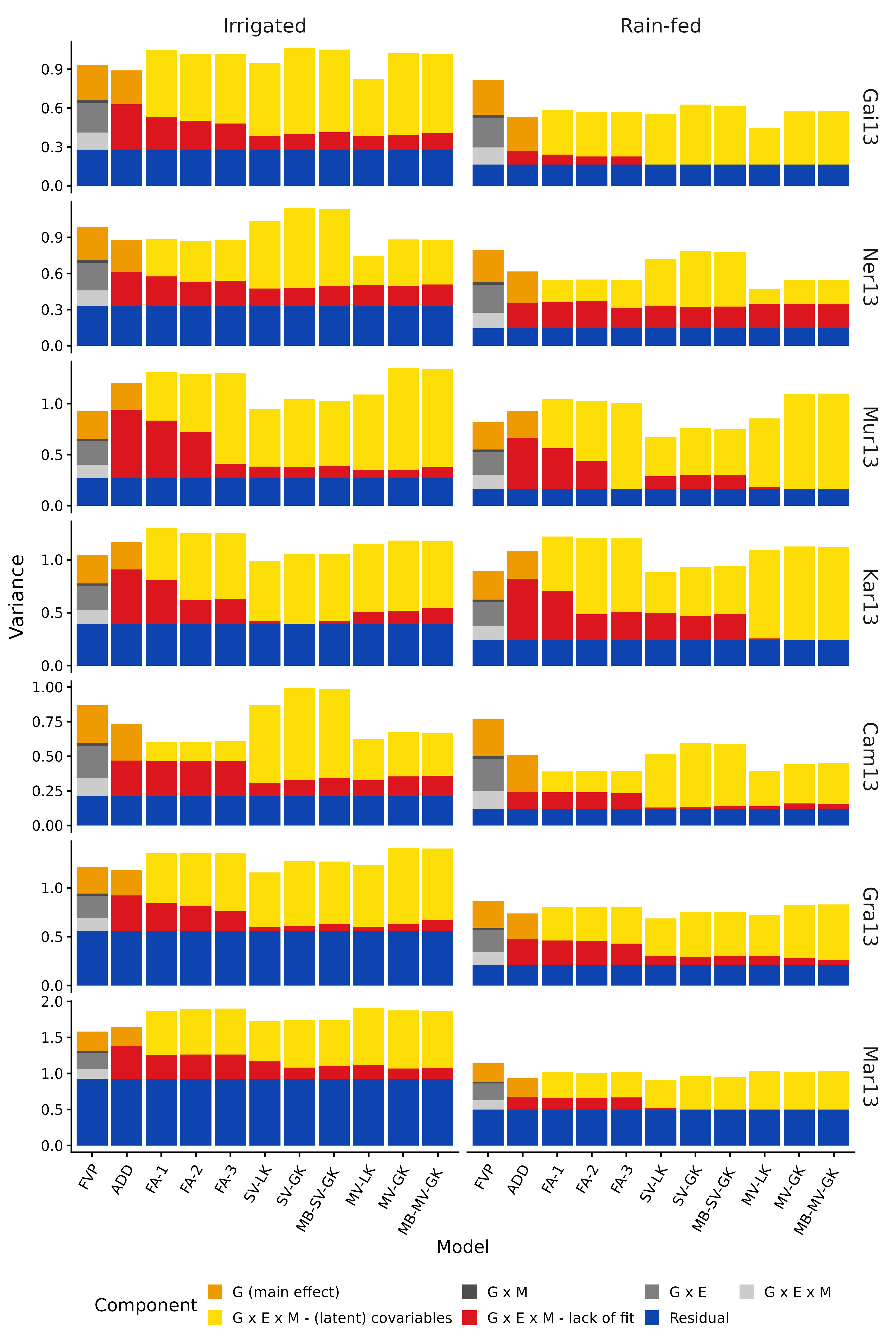}
	\caption{Continued. Partitioning of G$\times$E$\times$M variance into structured, lack of fit, and residual effects for each environment in the DROPS dataset.
	\\[5pt]
	\textbf{Alt text:} Bar charts showing the amounts of variance explained by the different effects in the variance partitioning models for the DROPS dataset for the remaining environments.}
	\label{fig:DROPS_LOF_EXTRA_B}
\end{figure}

\begin{figure}[h]
	\includegraphics[width=0.85\textwidth]{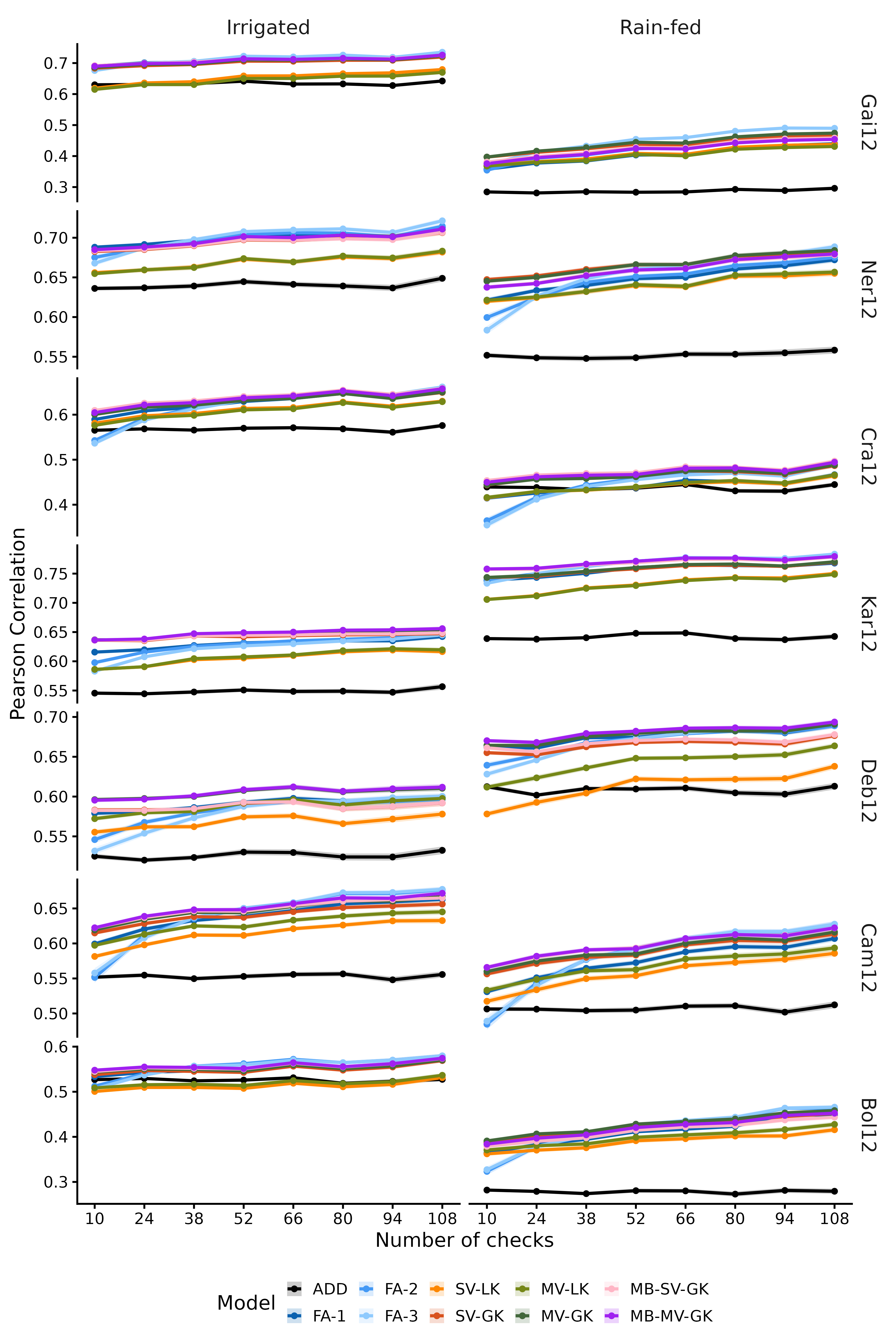}
	\caption{Pearson correlation between BLUPs from the LMMs and centered test-set phenotypes for the DROPS dataset for each environment. Shaded areas represent standard errors.
	\\[5pt]
	\textbf{Alt text:} Line graphs showing the genomic prediction accuracies (Pearson correlation) obtained using the different models and levels of sparsity for the first seven environments of the DROPS dataset.}
	\label{fig:DROPS_CORR_EXTRA_A}
\end{figure}
\begin{figure}[h]
	\includegraphics[width=0.85\textwidth]{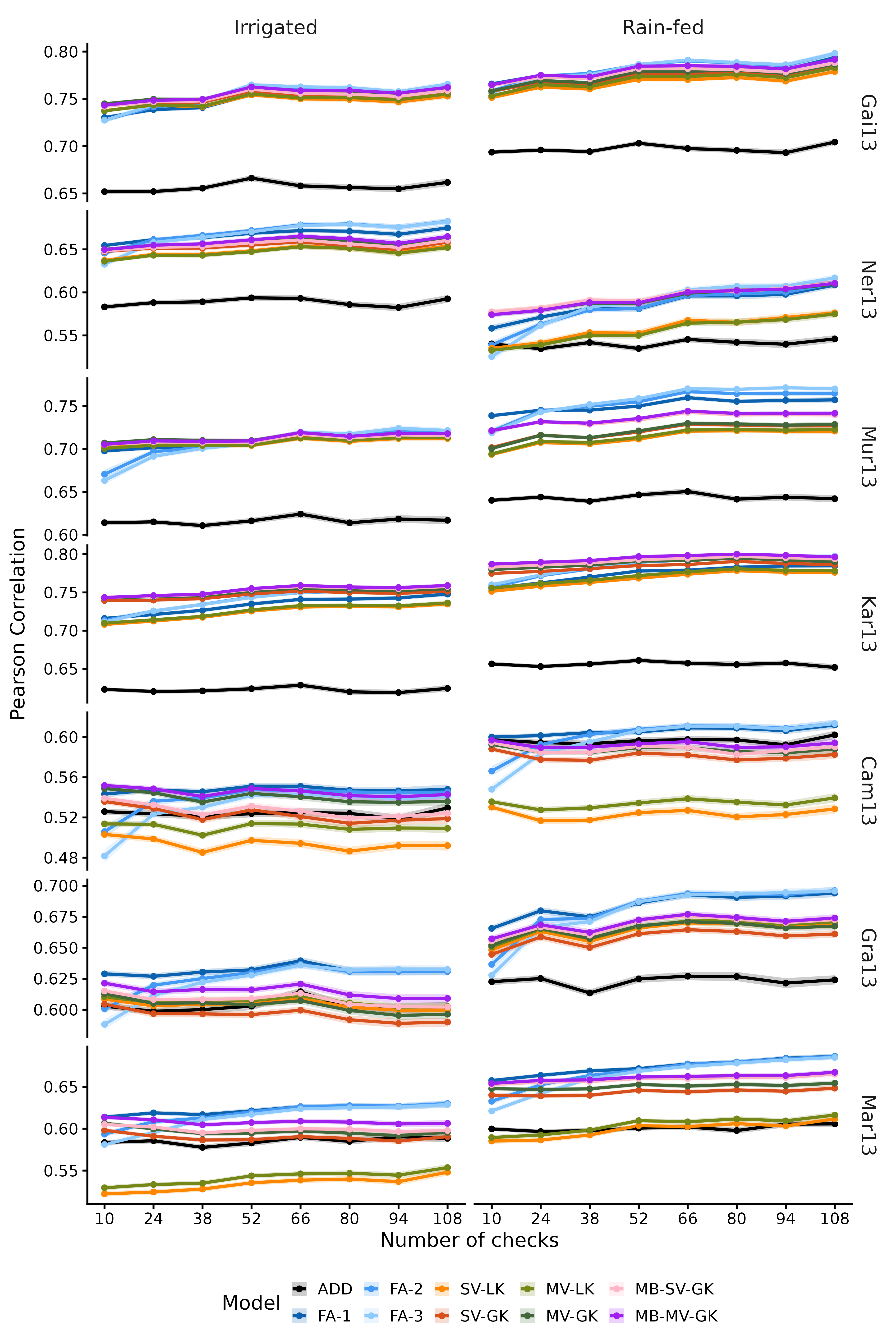}
	\caption{Continued. Pearson correlation between BLUPs from the LMMs and centered test-set phenotypes for the DROPS dataset for each environment. Shaded areas represent standard errors.
	\\[5pt]
	\textbf{Alt text:} Line graphs showing the genomic prediction accuracies (Pearson correlation) obtained using the different models and levels of sparsity for the remaining environments of the DROPS dataset.}
	\label{fig:DROPS_CORR_EXTRA_B}
\end{figure}
\begin{figure}[h]
	\includegraphics[width=0.85\textwidth]{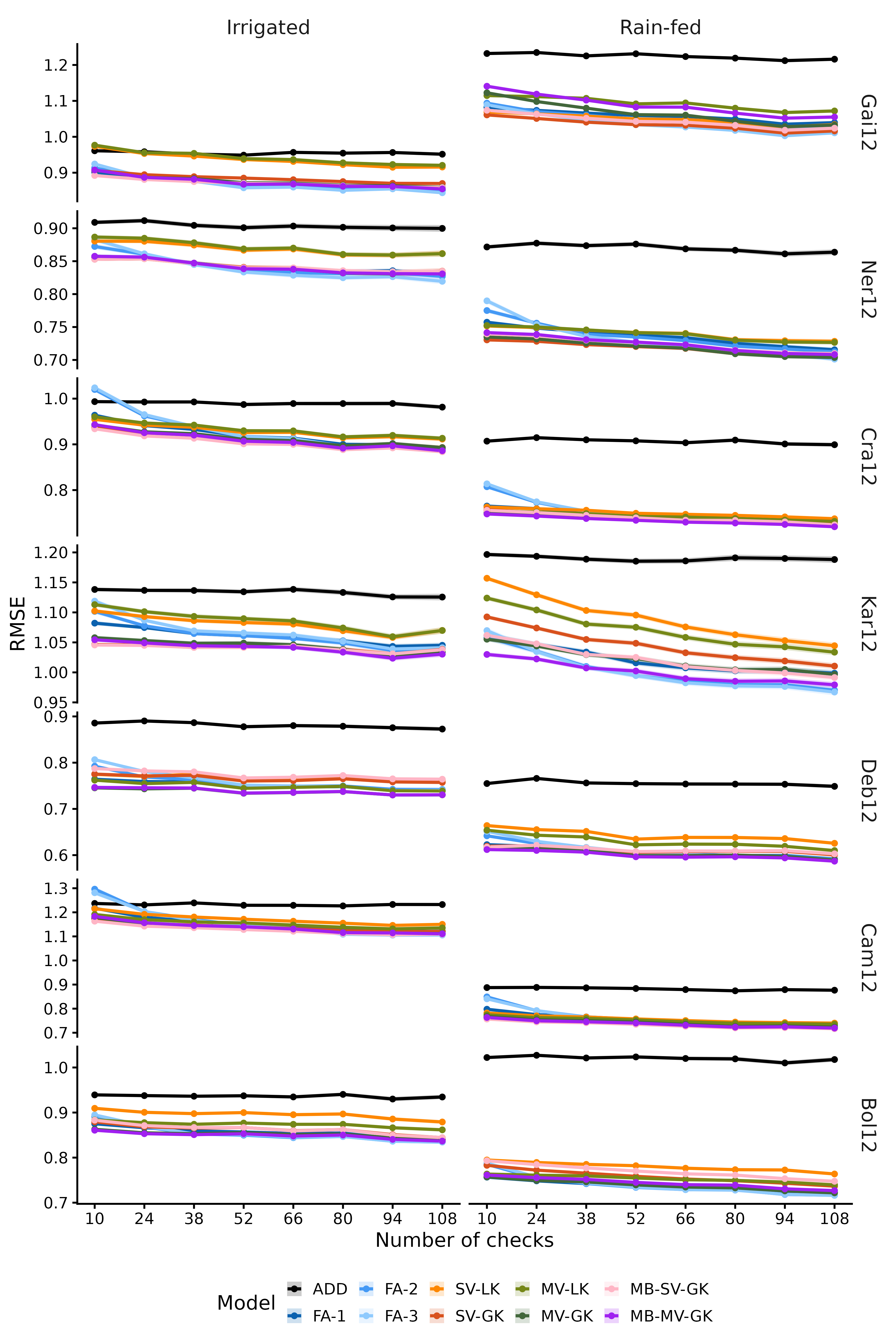}
	\caption{RMSE between BLUPs from the LMMs and centered test-set phenotypes for the DROPS dataset for each environment. Shaded areas represent standard errors.
	\\[5pt]
	\textbf{Alt text:} Line graphs showing the genomic prediction accuracies (RMSE) obtained using the different models and levels of sparsity for the first seven environments of the DROPS dataset.}
	\label{fig:DROPS_RMSE_EXTRA_A}
\end{figure}
\begin{figure}[h]
	\includegraphics[width=0.85\textwidth]{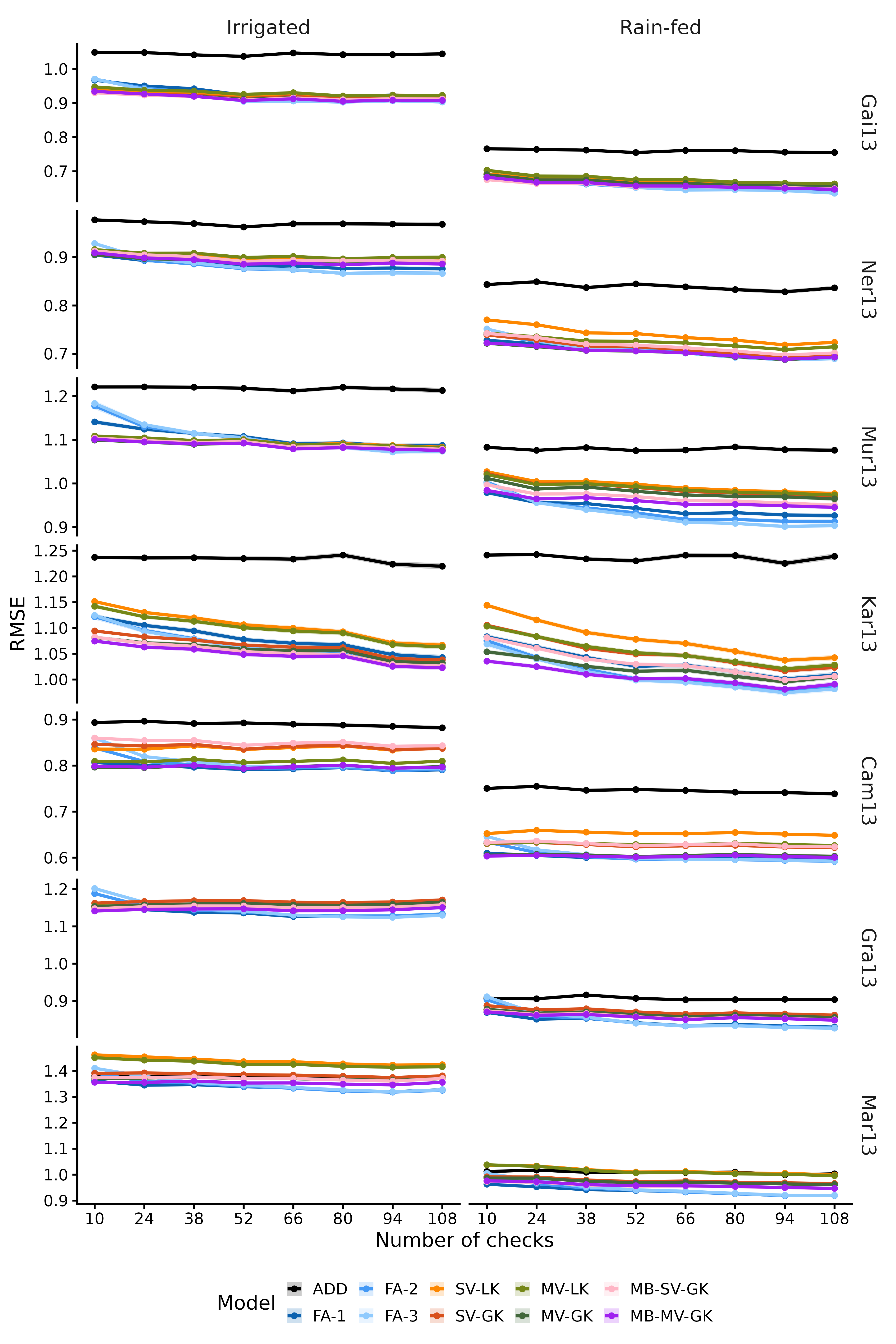}
	\caption{Continued. RMSE between BLUPs from the LMMs and centered test-set phenotypes for the DROPS dataset for each environment. Shaded areas represent standard errors.
	\\[5pt]
	\textbf{Alt text:} Line graphs showing the genomic prediction accuracies (RMSE) obtained using the different models and levels of sparsity for the remaining environments of the DROPS dataset.}
	\label{fig:DROPS_RMSE_EXTRA_B}
\end{figure}
\begin{figure}[h]
	\includegraphics[width=0.85\textwidth]{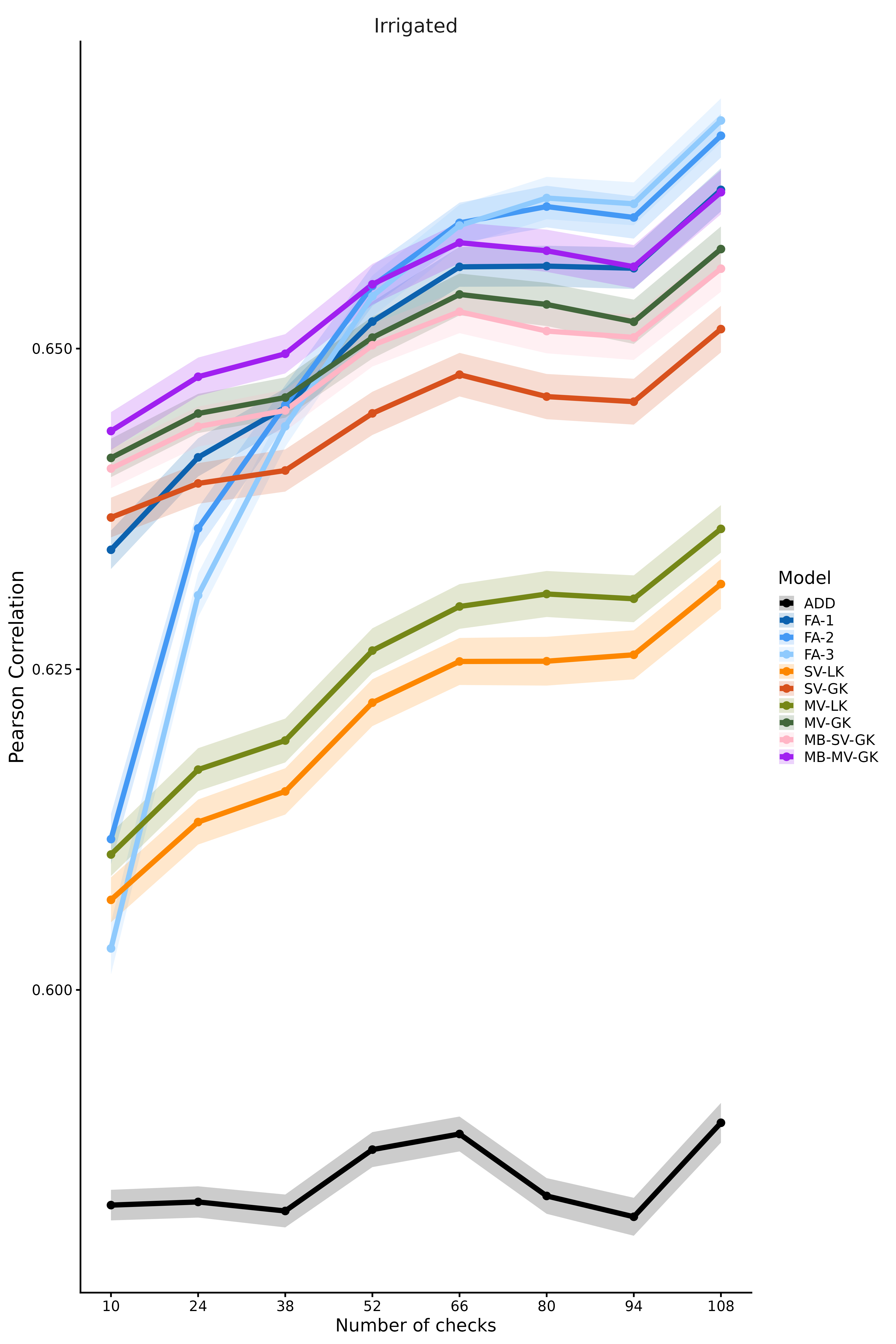}
	\caption{An enlarged version of the Pearson correlation, irrigated panel of Figure \ref{fig:DROPS_CV2}.
		\\[5pt]
		\textbf{Alt text:} An enlarged version of the Pearson correlation, irrigated panel of Figure \ref{fig:DROPS_CV2}.}
	\label{fig:DROPS_CV2_PCW}
\end{figure}
\begin{figure}[h]
	\includegraphics[width=0.85\textwidth]{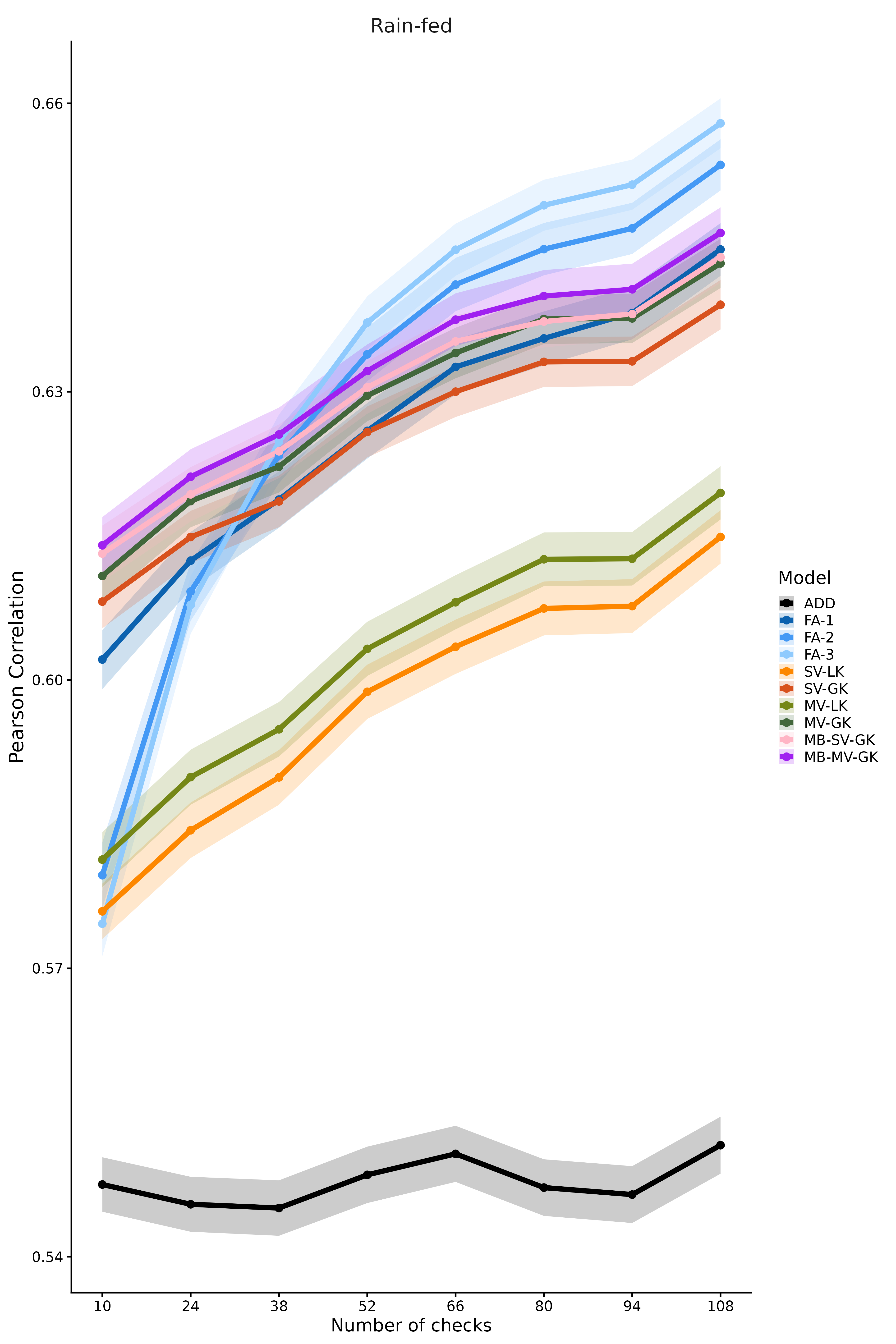}
	\caption{An enlarged version of the Pearson correlation, rain-fed panel of Figure \ref{fig:DROPS_CV2}.
		\\[5pt]
		\textbf{Alt text:} An enlarged version of the Pearson correlation, rain-fed panel of Figure \ref{fig:DROPS_CV2}.}
	\label{fig:DROPS_CV2_PCR}
\end{figure}
\begin{figure}[h]
	\includegraphics[width=0.85\textwidth]{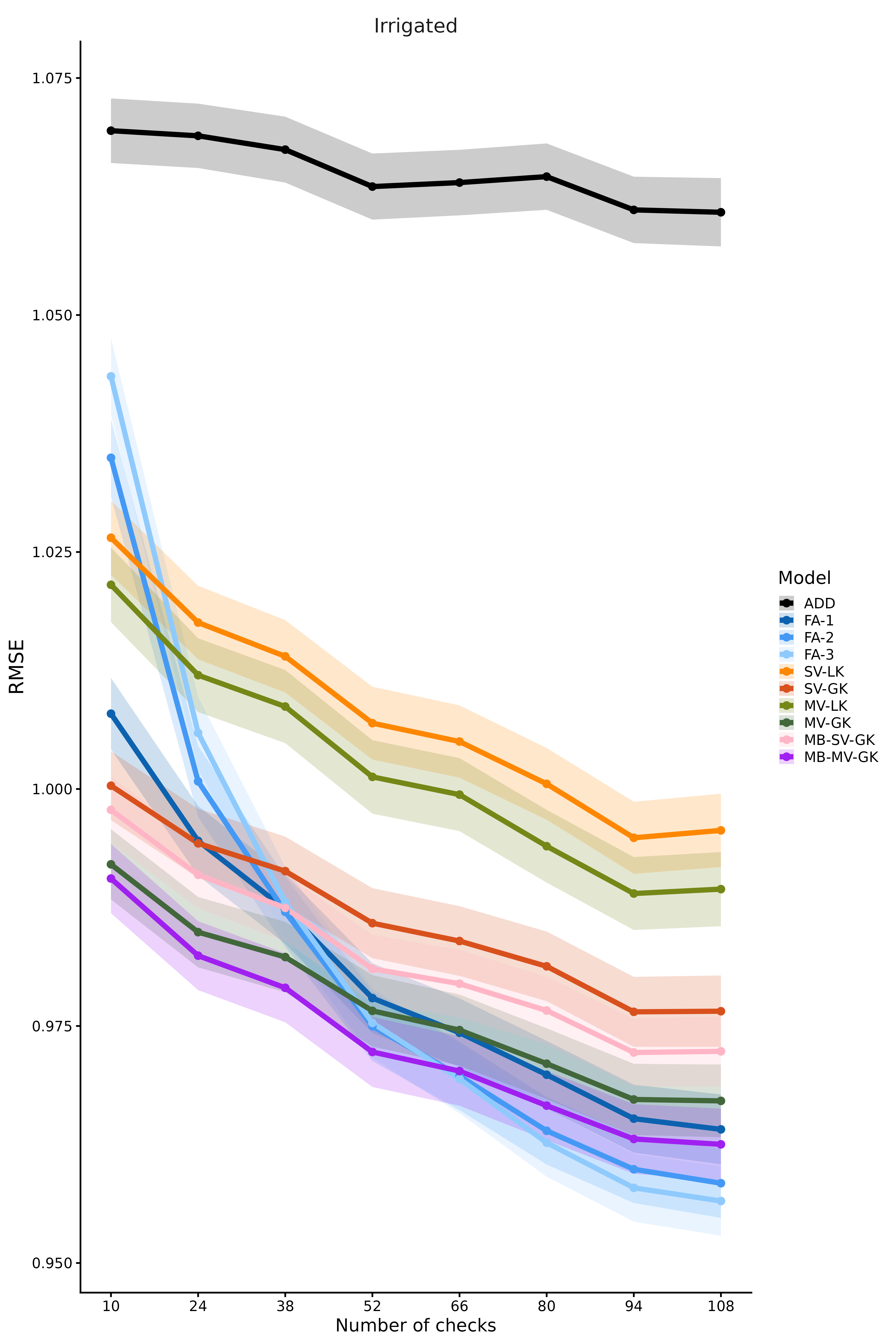}
	\caption{An enlarged version of the RMSE, irrigated panel of Figure \ref{fig:DROPS_CV2}.
		\\[5pt]
		\textbf{Alt text:} An enlarged version of the RMSE, irrigated panel of Figure \ref{fig:DROPS_CV2}.}
	\label{fig:DROPS_CV2_RMSEW}
\end{figure}
\begin{figure}[h]
	\includegraphics[width=0.85\textwidth]{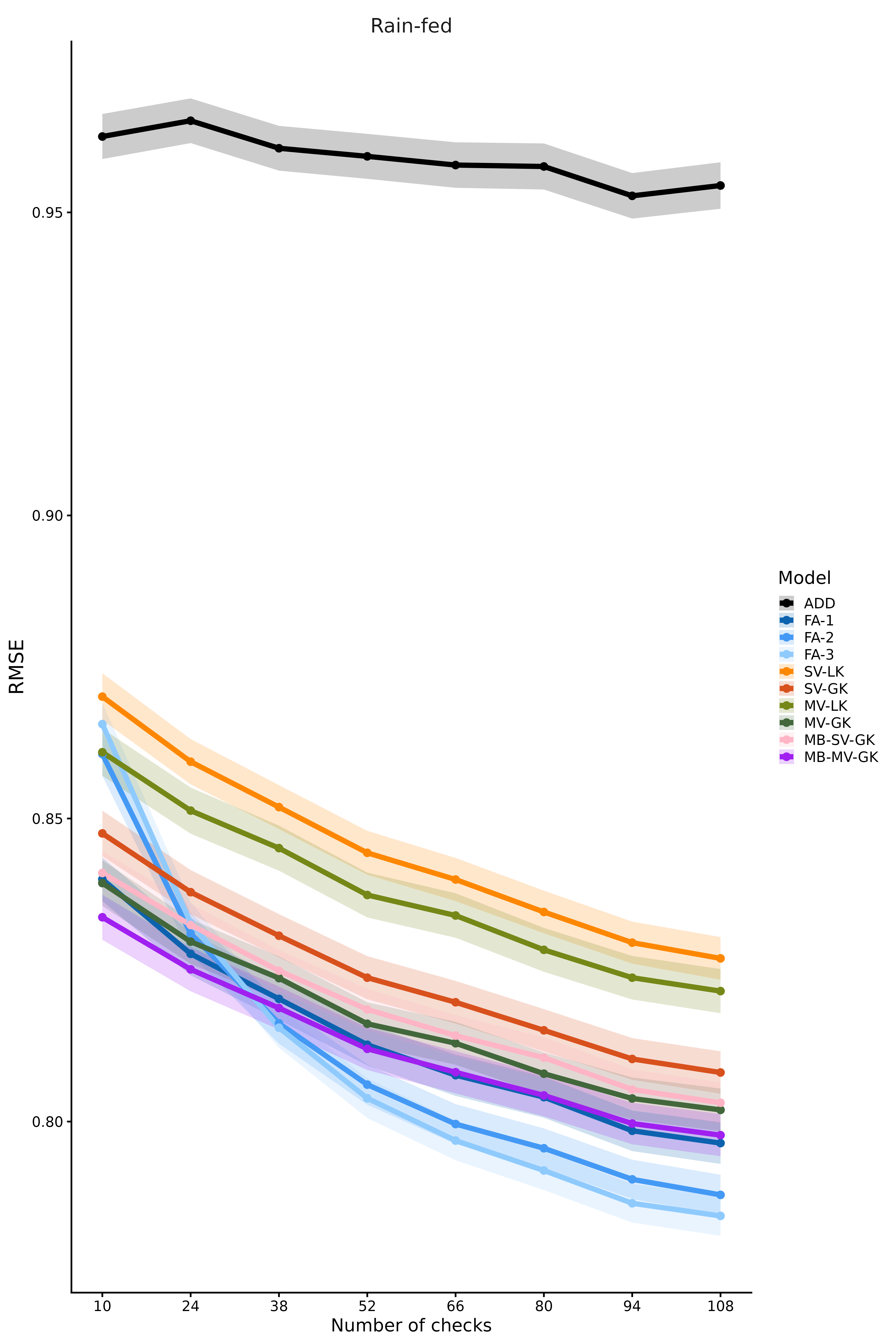}
	\caption{An enlarged version of the RMSE, rain-fed panel of Figure \ref{fig:DROPS_CV2}.
		\\[5pt]
		\textbf{Alt text:} An enlarged version of the RMSE, rain-fed panel of Figure \ref{fig:DROPS_CV2}.}
	\label{fig:DROPS_CV2_RMSER}
\end{figure}
\end{document}